\documentclass[fleqn,usenatbib]{mnras}
\usepackage{newtxtext,newtxmath}

\usepackage[T1]{fontenc}

\usepackage{bm}
\usepackage{xspace}
\usepackage{enumitem}
\usepackage{graphicx}
\usepackage{color}
\usepackage{comment}
\usepackage{amsmath}

\usepackage{array}
\usepackage{bm}
\usepackage[T1]{fontenc}
\usepackage{ae,aecompl}
\usepackage{graphicx}
\usepackage{float}
\usepackage{threeparttable}
\usepackage{comment}
\restylefloat{figure}
\usepackage{url}
\usepackage{hyperref}
\usepackage[labelfont=bf,labelsep=period]{caption}

\DeclareRobustCommand{\VAN}[3]{#2}
\let\VANthebibliography\thebibliography
\def\thebibliography{\DeclareRobustCommand{\VAN}[3]{##3}\VANthebibliography}

\usepackage{graphicx}	
\usepackage{amsmath}	

\usepackage{bm}
\usepackage{xspace}
\usepackage{enumitem}
\usepackage{color}
\usepackage{comment}

\providecommand{\dodoi}[1]{doi:~\href{http://doi.org/#1}{\nolinkurl{#1}}}
\providecommand{\doarXiv}[1]{arXiv:~\href{https://arxiv.org/abs/#1}{\nolinkurl{#1}}}

\usepackage{hyperref}
\usepackage{academicons}
\usepackage{xcolor}
\usepackage{orcidlink}

\title[Modeling TDE Optical/UV light curves]{Optical/UV Emission in the Tidal Disruption Event ASASSN-14li: Implications of Disc Modeling}

\author[Wen et al.]{
Sixiang Wen,\orcidlink{0000-0002-0934-2686}$^{1,2}$\thanks{E-mail:S.Wen@astro.ru.nl}
Peter G.~Jonker,\orcidlink{0000-0001-5679-0695}$^{1,3}$
Nicholas C. Stone,\orcidlink{0000-0002-4337-9458}$^{4}$
Sjoert Van Velzen,\orcidlink{0000-0002-3859-8074}$^{5}$
and Ann I. Zabludoff, \orcidlink{0000-0001-6047-8469}$^{2}$
\\
$^{1}$Department of Astrophysics/IMAPP, Radboud University, P.O.~Box 9010, 6500 GL Nijmegen, The Netherlands\\
$^{2}$University of Arizona, 933 N. Cherry Ave., Tucson, AZ  85721\\
$^{3}$SRON, Netherlands Institute for Space Research, Niels Bohrweg 4, 2333 CA, Leiden, The Netherlands\\
$^{4}$Racah Institute of Physics, The Hebrew University, Jerusalem, 91904, Israel\\
$^{5}$Leiden Observatory, Leiden University, Postbus 9513, 2300 RA, Leiden, The Netherlands
}

\date{Accepted 2023 March 30. Received 2023 February 18; in original form 2022 October 24}

\pubyear{2023}

\begin{document}
\label{firstpage}
\pagerange{\pageref{firstpage}--\pageref{lastpage}}
\maketitle

\begin{abstract}
We predict late-time optical/UV emission from tidal disruption events (TDEs) from our slim accretion disc model \citep{Wen20} and explore the impact of the black hole mass $M_\bullet$, black hole spin $a_\bullet$, and accretion disc size. We use these synthetic spectra to successfully fit the multi-band \emph{Swift} observations of ASASSN-14li at >350 days, setting only the host galaxy extinction and outer disc radius as free parameters and employing the $M_\bullet$, $a_\bullet$, disc inclination, and disc accretion rates derived from fitting 10 epochs of ASASSN-14li's X-ray spectra with the slim disc. To address the nature of the \emph{early}-time optical/UV emission, we consider two models: shock dissipation and reprocessing. We find that (1) the predicted late-time optical/UV colour (e.g., $u-w2$) is insensitive to black hole and disc parameters unless the disc spreads quickly; (2) a starburst galaxy extinction model is required to fit the data, consistent with ASASSN-14li's post-starburst host; (3) surprisingly, the outer disc radius is $\approx$2$\times$ the tidal radius and $\sim$constant at late times, showing that viscous spreading is slow or non-existent; (4) the shock model can be self-consistent if $M_\bullet \lesssim 10^{6.75}$M$_\odot$, i.e., on the low end of ASASSN-14li's $M_\bullet$ range ($10^{6.5-7.1}$M$_\odot$; 1$\sigma$ CL); larger black hole masses require disruption of an unrealistically massive progenitor star; (5) the gas mass needed for reprocessing, whether by a quasi-static or an outflowing layer, can be $<0.5$M$_\odot$, consistent with a (plausible) disruption of a solar-mass star.
\end{abstract}



\section{Introduction}
\label{int}

Tidal disruption events (TDEs) happen when a star approaches a supermassive black hole (SMBH; \citep{Hills75, Rees1988}). After the star is broken down by tidal forces, half of the debris remains bound and can be accreted by the SMBH, producing a strong electromagnetic flare. These flares frequently emit at optical \citep{vanVelzen+11, Gezari+12, Arcavi+14, Holoien2016a, vanVelzen+20}, near-ultraviolet \citep[NUV;][]{Gezari+06, Gezari+08}, and soft X-ray \citep{Bade+96, Greiner+00, Komossa+04, Saxton+21} wavelengths. 

The source of TDE optical/UV emission is still a matter of debate. There are two leading models for the early (power-law decaying) optical/UV light: one is emission from a reprocessing layer, which is powered by the X-rays and extreme UV emanating from the inner accretion disc \citep{Loeb+97, Guillochon+14, Metzger2016, Roth+16, RothKasen18, Dai+18, LuBonnerot19}, and the other is shock-powered emission, often assumed to involve an outer shock that forms at the intersection of the debris streams near their orbital apocenters \citep{Shiokawa+15,Piran+15}, but sometimes also involving shocks between tidal debris and a circularizing accretion flow \citep{BonnerotLu19,SteinbergStone22}. Based on these two different paradigms, two modeling suites, MOSFiT \citep{Mockler2018} and TDEMASS \citep{Taeho+2020}, respectively, have been used to constrain the masses of the SMBH and disrupted star by fitting the early observed optical/UV light curves. Both models yield SMBH mass constraints consistent with determinations from the black hole (BH) mass versus galaxy velocity dispersion relation \citep{Ferrarese2005,Kormendy2013,McConnell2013}.

In contrast to the optical/UV emission, there is more consensus on the origin of TDE thermal X-rays, which are widely thought to be powered by accretion and to arise from an inner disc \citep{Ulmer99}. The X-ray spectra of the TDEs ASASSN-14li, ASASSN-15oi, and J2150 \citep[][hereafter W20 and W21]{Wen20, Wen+21} are consistent with those predicted by a general relativistic slim accretion disc model \citep{Abramowicz1988}; additional X-ray-producing mechanisms are not obviously required, although at late times often the mass accretion rate can be well below the Eddington limit (the disc may experience a hard state change, e.g., \citep{Jonker+20}). From the X-ray spectral fitting of two TDEs, W20 found that the absorption parameter (i.e. the column density $N_{\rm H}$), declines to the level predicted for the host galaxy plus Milky Way contribution after several hundred days. This behavior supports a picture in which, at late times, a gaseous reprocessing layer\footnote{The absorption of disc X-rays by a time-variable absorbing column indicates that {\it some} reprocessing is occurring; whether or not it is the dominant contributor to the early-time optical/UV emission is an important question we will investigate later in this paper.} becomes increasingly optically thin, and the optical/UV emission, which has dimmed by $\gtrsim10\times$, is dominated by a disc component \citep{vanVelzen+19}. It is therefore important to test whether the observed optical/UV emission at late times can be successfully fit with a bare slim disc model, after allowing for host dust extinction and (potentially) a time-varying outer disc radius.
 
In this paper, we focus on ASASSN-14li's optical/UV emission, based on our successful fitting of its evolving X-ray spectrum (W20). ASASSN-14li was first discovered on 22 November 2014, in a post-starburst galaxy at $z=0.0206$ \citep{jose2014}. The optical/UV light curves initially declined quickly and then evolved more gradually after about 350 days \citep{Brown+17}. The slow late-time decay is consistent with disc-dominated emission \citep{vanVelzen+19} and can be well fit by a time-dependent model of a viscously spreading, general relativistic thin disc \citep[][hereafter MB20]{Mummery&Balbus20}. In MB20, the accretion rate at each observing epoch is determined by the time-dependent disc equations, and the SMBH mass is constrained to $1.45\times 10^6 < M_\bullet / M_\odot < 2.05 \times 10^6$ with a broad range of permitted BH spins.

The MB20 constraints on $M_\bullet$ and spin are only marginally consistent (at the $2\sigma$ confidence level (CL)) with the slim disc fitting results of W20. In this paper, we perform a detailed study of the late optical/UV emission of ASASSN-14li based on the results obtained through fitting the X-ray spectra with the slim disc. Unlike the MB20 model, the disc accretion rate here is a free parameter and is determined by the X-ray spectral fits. We check whether (1) the slim disc solution can match the late time optical/UV light curves; (2) dust extinction in the host galaxy is important for the observed optical/UV emission; and (3) the size of the accretion disc evolves, i.e., whether angular momentum transport may cause the disc to spread out with time. 

In addition to our modeling of the late-time light curves, we test two possible sources---a reprocessing debris layer or a shock arising from a self-intersecting debris stream---for the \emph{early-time} optical/UV emission of ASASSN-14li. In these tests, we calculate the minimum mass of the disrupted star that can produce the observed optical/UV emission, where the inferred disc luminosity,  $M_\bullet$, and dimensionless spin parameter $a_\bullet$ are derived from the X-ray spectral fits.

The paper is organized as follows.  We explore the general dependence of theoretical optical/UV light curves on $M_\bullet$, $a_\bullet$, and outer disc radius $R_{\rm out}$ in Section \ref{TOLC}. We describe the (archival) optical/UV emission data for ASASSN-14li and our reduction procedure in Section \ref{data}. In Section \ref{RandD}, we show the fit results of the slim disc model to the observed late-time optical/UV light curves. In Section \ref{earlylc}, we describe our tests of whether reprocessing or the shock paradigm can explain the bright early optical/UV emission. Finally, we summarize our conclusions in  \ref{Conclusions}.

\section{Theoretical Optical/UV Light Curves}
\label{TOLC}

In this section, we calculate the late-time optical/UV emission from the slim disc model of W21. We use a multi-colour black body model to calculate the local optical/UV emission and a ray-tracing code \citep{JP11}, which includes gravitational redshift, Doppler, and lensing effects self-consistently, to determine the monochromatic flux. 
As the disc formed after a TDE is finite, the disc size becomes an important factor to the optical/UV emission. Here, we limit the disc to radii $R<R_{\rm out}$, neglecting the small difference between our hard cutoff and the exponential 
cutoff used by MB20.

\begin{figure*}
\includegraphics[width=0.33\textwidth]{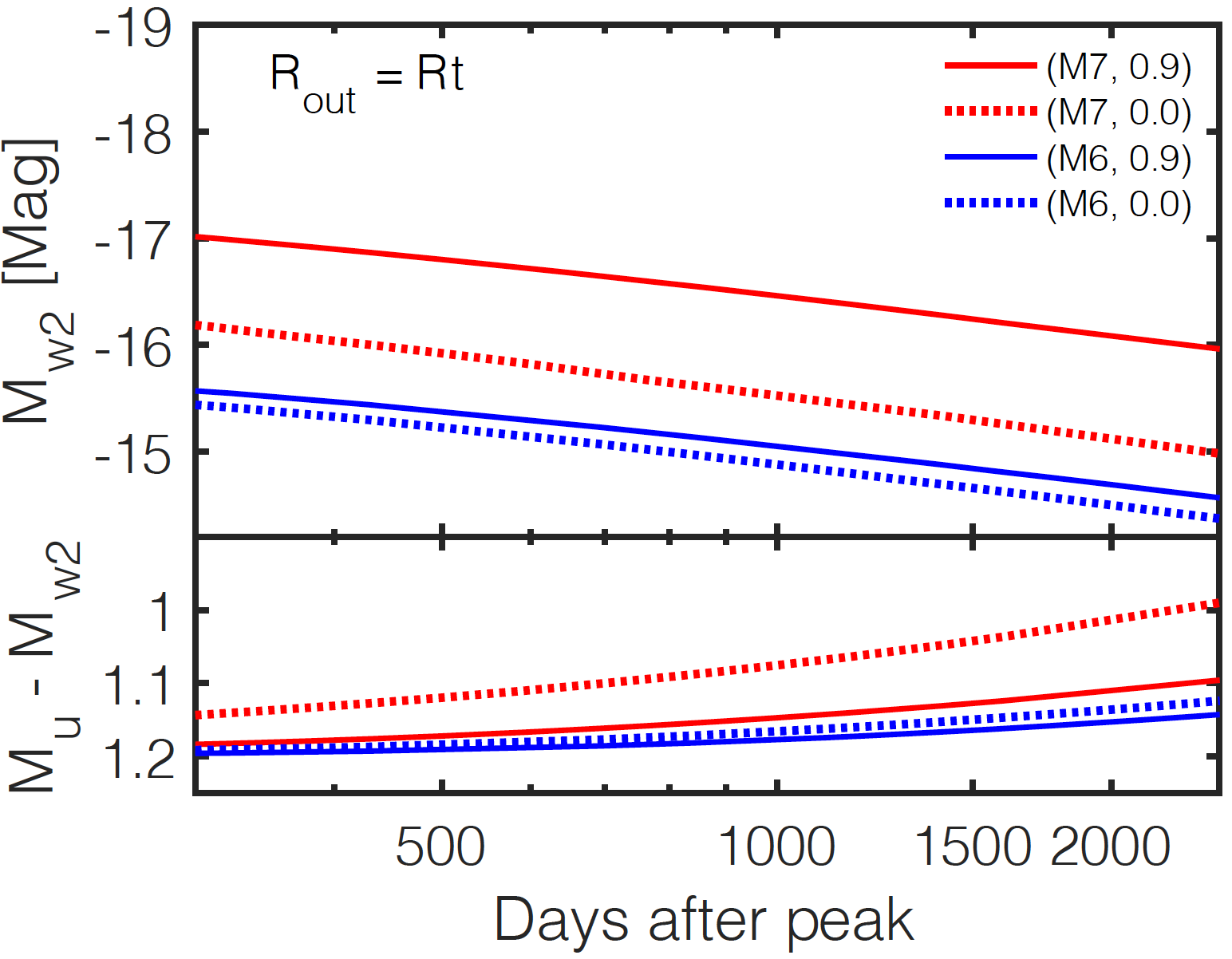}\hfill
\includegraphics[width=0.33\textwidth]{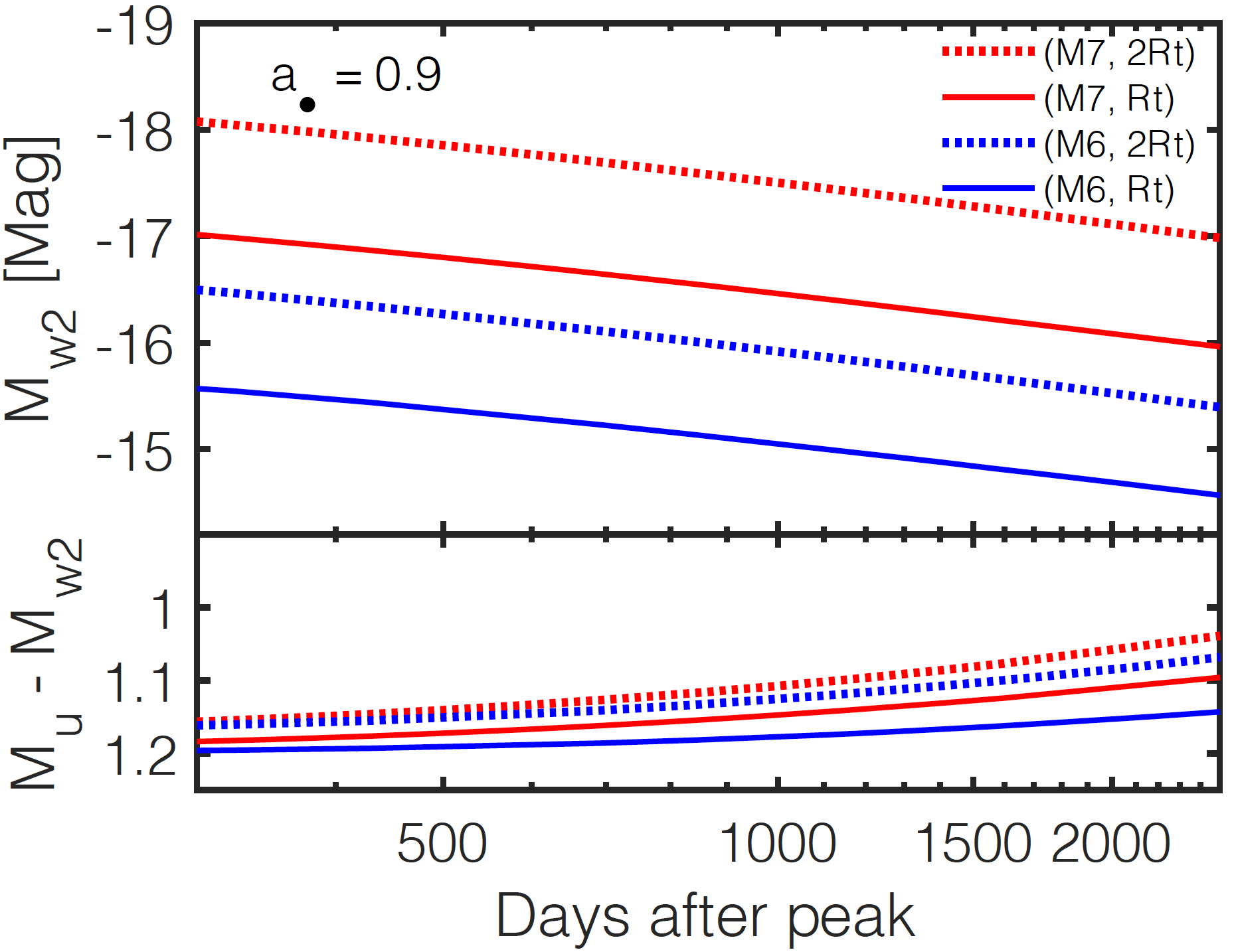}\hfill
\includegraphics[width=0.33\textwidth]{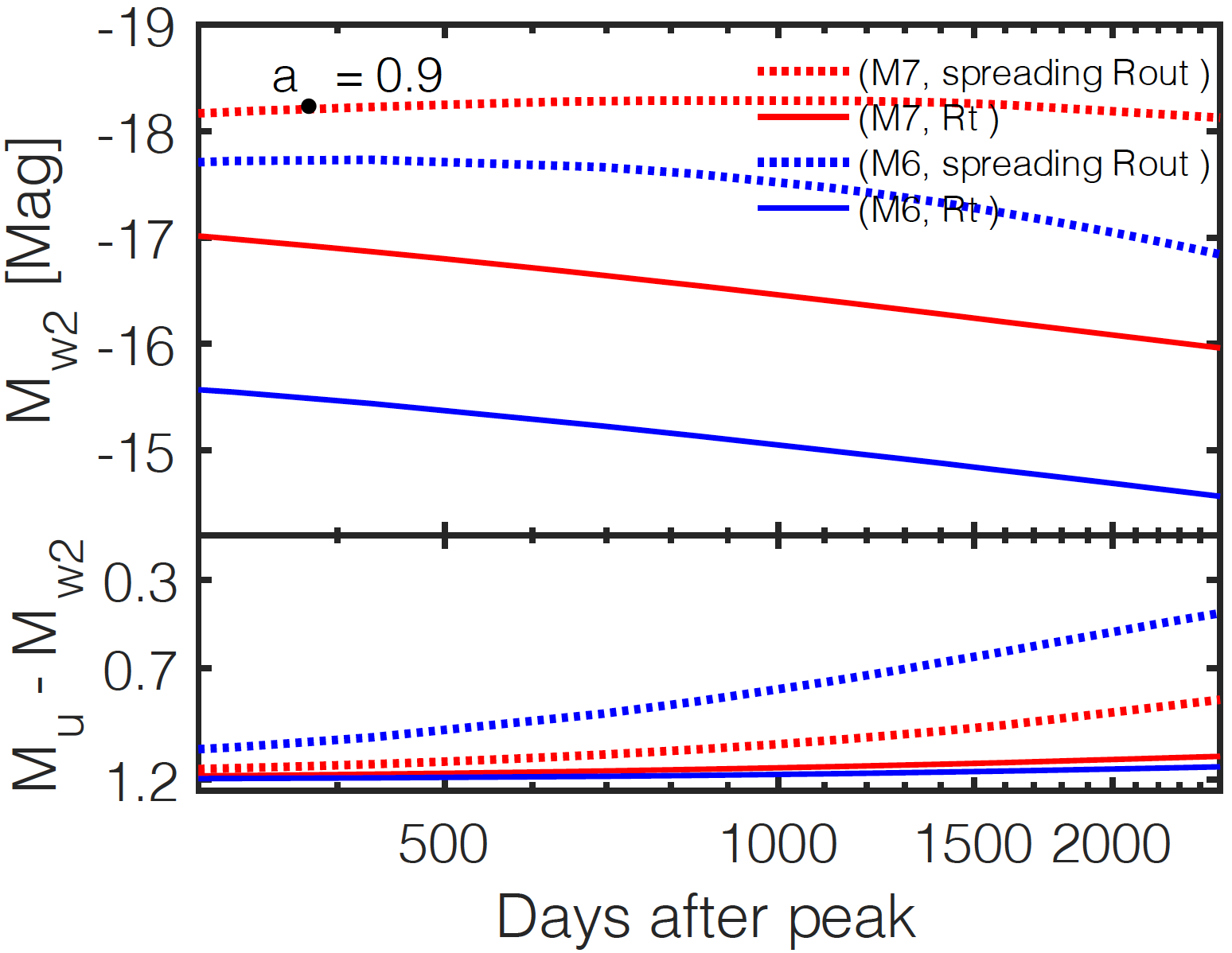}\hfill
\caption{Theoretical thermal slim disc optical/UV light curves for different  $M_\bullet$, $a_\bullet$, and outer disc radius $R_{\rm out}$. Here we assume  the viewing angle $\theta=45^\circ$, disc viscosity parameter  $\alpha=0.1$,  luminosity distance $d_L=10 ~{\rm pc}$, TDE penetration parameter $\beta=2$, and  progenitor star mass $M_\star=2M_\odot$ throughout. The left figure shows the light curves for different  $M_\bullet$ and $a_\bullet$, with the outer radius fixed at $R_{\rm out} = R_t$. The middle figure shows the light curves for different  $M_\bullet$ and $R_{\rm out}$, with $a_\bullet = 0.9$. 
The right figure considers a viscously spreading disc, but with an independent model for the viscous torque (see Appendix B). In all figures, we use the notation $\rm{MX}=10^X~M_\odot$. The upper sub-panels show the light curves in the $w2$ (1928~ \rm{\AA}) filter band, and the lower sub-panels show the difference in AB magnitude between the $u$ (3465~\rm{\AA}) and $w2$ 
bands. The figures show that (1) thermal disc UV emission is brighter for higher $M_\bullet$, as a higher $M_\bullet$ yields a larger physical disc size; (2) for $M_\bullet$, $a_\bullet$, and $R_{\rm out}$ there is a positive correlation between the value of the parameter and the normalisation of the light curve; (3) the spreading disc makes the light curves flatter than those with fixed $R_{\rm out}$;
(4) 
the $u-w2$ colour evolution is small (0.1--0.2 mag) over 
about 2000 days (except for the case of the viscously spreading disc) and it therefore insensitive to the parameters under consideration, e.g., $M_\bullet$, $a_\bullet$ and $R_{\rm out}$ and $\dot m$ (the $\dot m$ dependence is implicit as different epochs have different accretion rates). Furthermore, such a colour evolution will be difficult to measure given the typical measurement uncertainties, including those incurred due to subtracting the host galaxy light and uncertainties in the extinction.}
\label{Fig:thlc}
\end{figure*}

We explore the impact of $M_\bullet$, $a_\bullet$, and $R_{\rm out}$ on theoretical optical/UV light curves in Figure~\ref{Fig:thlc}. In this analysis, we estimate the disc accretion rate from hydrodynamic simulations of the disruption process \citep{GuillochonRamirezRuiz13}, which can be determined from the time after peak, $M_\bullet$, the penetration parameter $\beta$, the progenitor star mass $M_\star$, and  polytropic index (here we consider the case of $\gamma=4/3$). The left plot shows the light curves for different values of $M_\bullet$ and $a_\bullet$; we fix the disc outer radius to $R_{\rm out}= 2R_t/\beta$ ($R_t$ is tidal radius). A larger $M_\bullet$ always results in a brighter optical/UV source, because  
larger SMBHs produce accretion discs with lower effective temperatures, moving the optical/UV bands up the Rayleigh-Jeans tail and closer to the peak of the multi-colour blackbody spectrum \citep{LodatoRossi11}.  The effect of spin on optical/UV emission is strong for a high mass BH, but weak for a low mass BH. This behavior arises because all the optical/UV emission is generated by the disc relatively close to the BH ($R_{\rm out} \lesssim 20R_g$, $R_g=GM_\bullet/c^2$) for discs associated with larger $M_\bullet$, while for a low mass BH, most of the optical/UV emission is generated in a region relatively far from the BH. The colour $u - w2$, based on bands from the \textit{Neil Gehrels Swift Observatory} \citep{Gehrels+04}, is roughly 1.1 mag, and it is not very sensitive to $M_\bullet$ and $a_\bullet$. 

The middle plot shows the light curves for different combinations of $M_\bullet$ and $R_{\rm out}$. Here, we fix $a_\bullet= 0.9$. As the disc size increases by a factor of 2, the optical/UV emission brightens by about 1 mag. The larger disc radius increases the disc area by a factor of four, while the UV flux increases\footnote{The outer disc is cooler than the inner disc. As the inner hotter disc contributes more optical/UV photons per area than the outer disc, the integrated flux increases more slowly than the disc size as $R_{\rm out}$ grows.} by a factor of $\sim2.5$. As a result, the cooler larger disc also contributes a significant part to the optical/UV emission. Again, the colour $u - w2$ is roughly 1.1 mag, and the value and its evolution are not very sensitive to disc size. Since the disc UV emission follows the Rayleigh–Jeans law, given an accretion rate $\dot m_t \propto t^{-n}$ (here, n=1.53), the UV light curve decays as,
\begin{equation}
\rm{M}_{\rm UV} \propto \frac{2.5n}{4}\log_{10}(t).
\label{Mag}
\end{equation}
The left and middle figures show good agreement with the linear relationship between magnitude and time. However, we note that the above relationship is only valid for a disc with constant $R_{\rm out}$; if the disc can viscously spread \citep{Cannizzo+90}, then the light curves will decay more slowly with time.

The right panel of Figure~\ref{Fig:thlc} shows the effect of viscous disc spreading.
As the TDE disc is finite, the accreted debris transfers part of its angular momentum outward through the viscous torque, making the disc spread outwards over time \citep{Cannizzo+90, Mummery&Balbus20}. 
Because of the well-known viscous instability of realistic parametrizations of angular momentum transport in hot accretion flows \citep{LightmanEardley84}, it is not possible to build a self-consistent 1D time-dependent model for a spreading disc dominated by radiation pressure.  We instead develop a toy model to estimate the $R_{\rm out}$ of the spreading disc (see Appendix \ref{app:spread}). In this model, we ignore the specific nature of the viscous torque, and only require conservation of mass and angular momentum. We use Eq. \ref{app:rout} to evaluate the spreading $R_{\rm out}$ at different time. When plotting the figure, we adopt $\Gamma=-3/2$ and use the mass fallback rate to estimate the mass accreted ($\Delta M$) by the BH. 

The UV light curve of spreading disc is at least 1 mag brighter (after 300 days) than the UV light curve of the disc with fixed $R_{\rm out}$, and its decay is much flatter. The disc radius $R_{\rm out}$ extends by a factor of 5 and 13 in 2000 days, for case of M7 and M6, respectively.  We see the colour also undergoes significant changes (0.5 mag). 
The precise nature of viscous spreading can place additional constraints on the $M_\bullet$ and $a_\bullet$ estimation, e.g., the UV light curve predicted from the X-ray spectral fits (without a spreading disc) would decay faster than the observed UV light curve.

\section{Observed Optical/UV Light Curves for ASASSN-14\lowercase{li}}
\label{data}

The TDE ASASSN-14li \citep{Holoien2016a} has been monitored for over five years using \emph{Swift}, with the latest observation obtained on December 2, 2020. We reduced all UVOT  \citep{Roming+05} data using the latest calibration files (20201215) and software (heasoft 6.29). Multiple sub-exposures within with the same observation ID are combined in the image plane before applying aperture photometry. The aperture radius is fixed to 5" and a curve of growth aperture correction is applied.  

In the latest epochs the UV flux still exceeds the pre-flare baseline, as inferred from the broadband SED of the host galaxy \citep{vanVelzen+20}, by about 0.5 mag. About 400 days after the first detection \citep{Brown+17}, the host-subtracted UV light curves flatten to a near-constant plateau \citep{vanVelzen+19}. In this work we focus on this late-time plateau. We apply our models to the difference photometry (i.e., after baseline subtraction) and apply a correction for Galactic extinction.

\section{Modeling ASASSN-14\lowercase{li's} Late-Time 
Emission}
\label{RandD}

In this section, we fit the late-time UV light curves of ASASSN-14li. The UV light curves are corrected for extinction by neutral gas and dust located in the Milky Way in the line of sight towards ASASSN-14li, but not for any contribution to the extinction from gas and dust located in the host galaxy. As a result, we explore the effect of gas and dust extinction from the host galaxy taking it as a free parameter in our modeling. For simplicity, we assume the same extinction across all late-time epochs. We consider 7 kinds of extinction curves from {\sc pysynphot} \citep{Lim+15}. They are {\sc lmc30dor} \citep{Gordon+03}, {\sc lmcavg} \citep{Gordon+03}, {\sc mwdense} \citep{Cardelli+89}, {\sc mwavg} \citep{Cardelli+89}, {\sc mwrv21} \citep{Cardelli+89}, {\sc mwrv4} \citep{Cardelli+89}, and {\sc xgalsb} \citep{Calzetti+00}. These models describe three kinds of dust grains environment, e.g. general Milky Way extinction environment \citep{Gordon+03},  diffuse and dense interstellar medium \citep{Cardelli+89}, dust environment in starburst galaxies \citep{Calzetti+00}. The extinction effect for all the 7 models can be described by a single parameter colour excess, $\rm{E_{B-V}}$, since the different $R_V$ values are taken into account in the different models.

We refit the X-ray spectra of ASASSN-14li using the slim disc model of W21 (see Appendix \ref{app:decay}, as well as \citep{Wen+22}). 
From the X-ray spectral fits, for a pair of $M_\bullet$ and $a_\bullet$, we get the best-fit accretion rate at each epoch, as well as the disc inclination. 
To calculate the UV emission at different epochs, we estimate the accretion rate at each epoch, using the accretion rate decay law obtained through the X-ray spectral fits, and set the disc outer radius $R_{\rm out}$ as a free parameter. However, as the disc may viscously spread with time, we bin the light curves in several bins and allow the fit to the data in each of the different bins to adopt a different $R_{\rm out}$. We will explore the effect of binning method on the fitted $R_{\rm out}$ in more details in section \ref{sec:host-ext} (see Table \ref{Tab:uv}). The free parameters are  $\rm{E_{B-V}}$ and $R_{\rm out,i}$, where i denotes different bins number.

In order to compare the predictions with the {\it Swift} observations, we simply compute the magnitude at the central wavelength of each UV filter band. For the four {\it Swift} bands of interest ($w2$, $m2$, $w1$, $u$) these are\footnote{\url{http://www.swift.ac.uk/analysis/uvot/filters.php}}: $w2=1928$~\AA, $m2=2246$~\AA, $w1=2600$~\AA, and $u=3465$~\AA. We fit the observations using a fit-function comprised of the slim disc model attenuated by extinction. We use the {\sc powell} method \citep{Press2002} to search for the best fit $\chi^2$. We quote the $1\sigma$ error of each parameter (single parameter) by adopting $\Delta \chi^2=1$.

\subsection{Successful Fits with a Bare Accretion Disc}

\begin{figure*}
\includegraphics[height=.55\textheight]{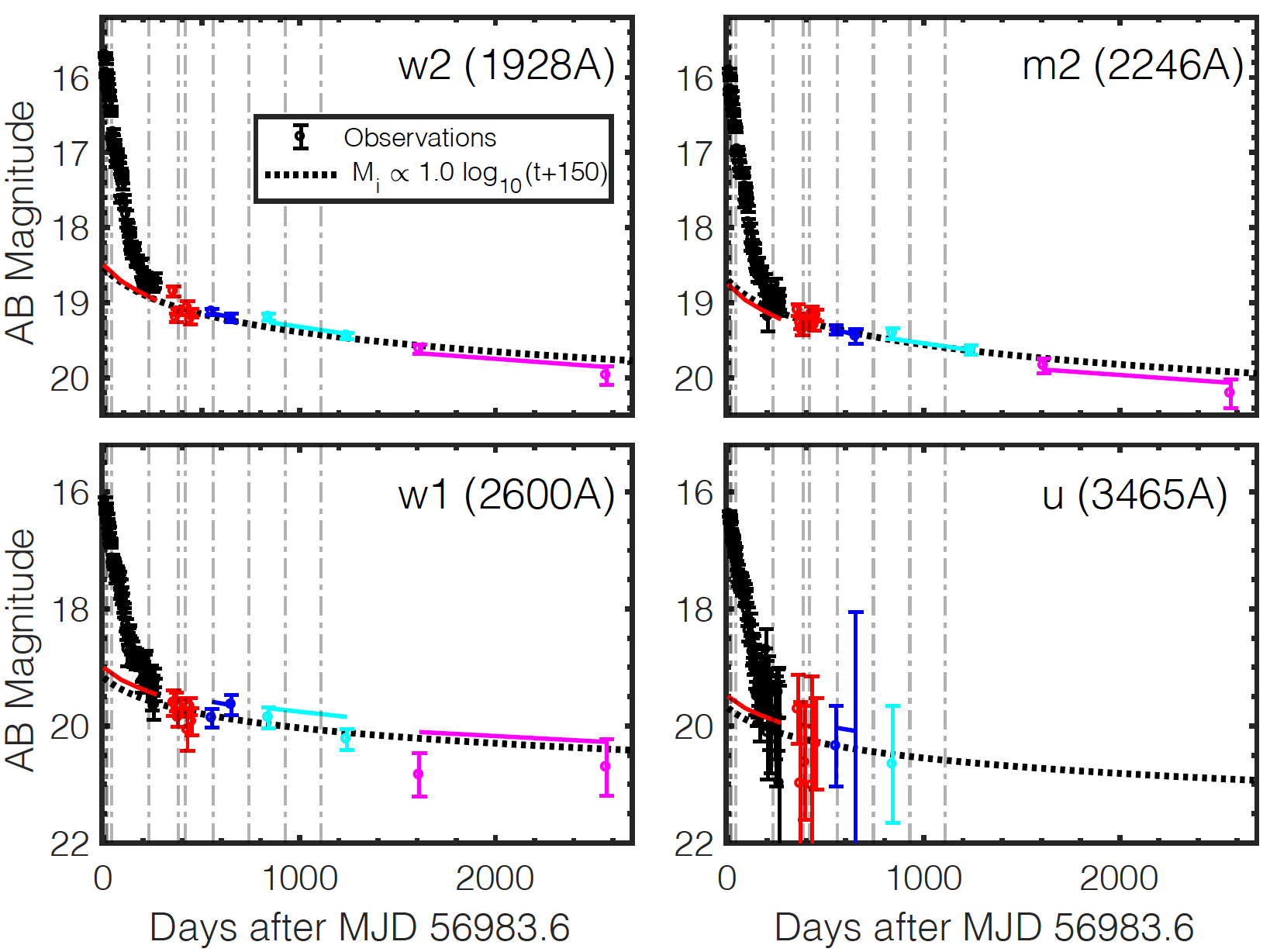}
\caption{Best fits to the late-time UV light curves in four different optical/UV filter bands obtained through {\it Swift/UVOT} observations.
Before the fit, the observed magnitudes have been corrected for extinction due to our own Galaxy, but not for any host extinction. Our approach assumes that there is no additional extinction associated with the TDE itself. The solid lines of 4 different colours denote the best-fit to the late-time (colour) data points using the slim disc model attenuated by a starburst extinction model \citep{Calzetti+00}. The black dotted lines show the best-fit to the data (excluding the black data points at early times) using a simple power-law model. The vertical gray dotted dashed lines show the dates of X-ray observations.  For the slim disc plus extinction fit, we fit the data averaged in the four groups denoted by different colours. This figure shows that the late-time UV emission of ASASSN-14li is consistent with the UV emission predicted from a slim disc, where most of the parameters of the slim disc are already determined from the X-ray spectral fits. The fitting results are listed in the $xgalsb_4$ row of Table~\ref{Tab:uv}.
}
\label{Fig:lateUV}
\end{figure*}

Figure \ref{Fig:lateUV} shows our best fit to the four observed UV light curves. Here we only fit the late-time epochs denoted with different colours (red, blue, cyan, and magenta). We will come back to the early-time (black) epochs in the next section. In this analysis, we fix $M_\bullet$, $a_\bullet$, and the disc inclination at the best-fit values obtained from our X-ray spectral fits, where $10^{+1}_{-7}\times 10^6M_\odot$, $a_\bullet=0.998_{-0.7}$ and $\theta=76^{+4}_{-74}$ ($0^\circ$ means face on). The accretion rate is also a parameter obtained from our X-ray spectral fits. Its evolution can be approximated with the following function: $\dot m_t= 231(t+150)^{-1.04\pm0.02}$. Here, $t$ is in days after the first detected by the All Sky Automated Search for Supernova (ASASSN) on MJD $56983.6$ in a post-starburst galaxy with $z=0.0206$ \citep{jose2014}, and $\dot m_t$ is the mass accretion rate in Eddington units with $\dot M_{\rm Edd} =1.37\times10^{15} M_\bullet/M_\odot ~\rm {kg~s}^{-1}$. We note that a different value of $M_\bullet$ and $a_\bullet$ will lead to a
somewhat different $\dot m_t$ equation (see Appendix \ref{app:decay})\footnote{Other choices of $M_\bullet$ and $a_\bullet$ lead, however, to very similar fitting results, e.g. $E_{\rm B-V} \sim 0.01$ and $R_{\rm out} =2 R_{\rm t}$ (see Fig. \ref{Fig:Rout}).}. 

We obtained 13 epochs of UV observations after $t=350$ d. In order to explore how well the model can fit an observation, we divided them into 4 groups as shown in Figure~\ref{Fig:lateUV}. For epochs within the same group, we assume that they have the same disc outer radius, which we parameterize as $R_1$, $R_2$, $R_3$ and $R_4$ for the 4 groups. In addition to the intrinsic UV emission from the slim disc model, we also employ one of the extinction models mentioned in the first paragraph of Section~\ref{RandD} to account for any host galaxy extinction. Therefore, the free parameters of this analysis are $R_1$, $R_2$, $R_3$, $R_4$ and $E_{B-V}$. We note that the fitting result of the parameters are not sensitive to the grouping method (see Table~\ref{Tab:uv}).

As can be seen from Figure~\ref{Fig:lateUV}, we successfully fit the four late time UV light curves simultaneously. Interestingly, the fit is only successful if we employ the starburst extinction model {\sc xgalsb} \citep{Calzetti+00} (we will investigate this in detail in Section~\ref{sec:host-ext}). This shows that the late-time UV emission is consistent with a disc spectrum, which is in line with the results of \cite{vanVelzen+19} and MB20. The reduced $\chi^2$ is $\chi^2/{\rm d.o.f}=69.38/43$, with the first bin deviating the most from the model (hence contributing most to the $\chi^2$). The discrepancy between the model and the data in the first bin may be explained if the UV light at that epoch is a combination of the UV slim disc emission and another source of UV emission.

We also fit the light curves with a power-law model $M_{\rm i}=k\log_{10}(t+150)+A_i$, where $M_{\rm i}$ denotes the magnitude of different bands and $A_i$ is the corresponding fitted intercept. We find a best fit with a reduced $\chi^2/{\rm d.o.f}=63.27/43$ and $M_{\rm i}\propto( 0.97\pm0.07) \log_{10} (t+150)$. The power-law model yields a better fit than that of the slim disc (both altered by an extinction model), likely because in the power-law fit the normalization of each light curve is determined separately, while the normalization of each light curve is associated with the disc spectrum and the extinction model for slim disc fit. The fitted power-law decays faster than that predicted by the slim disc model, i.e. $M\propto (0.66 \pm0.03) \log_{10} (t+150)$ inferred from the $\dot m_t=231(t+150)^{-1.04\pm0.02}$. We note that if we ignore the last two epochs, 
the fitted power law index is $0.77\pm0.15$, which is consistent with the predicted value of $0.66\pm0.03$ inferred from slim disc modeling \footnote{ As there is no X-ray spectrum available to calibrate the accretion rates after 1200 days, we compare the power law index from slim disc modeling with the power law index from the UV light curves fitting without the last two epochs.}. 

The power law index ($0.77\pm0.15$) obtained from fitting the late-time UV light curves indicates that, for a fixed $R_{\rm out}$ accretion disc, the disc accretion rate should decay as $1.2\pm0.2$ (see Eq.~\ref{Mag}) during the first $\sim 1200$ days. As the $R_{\rm out}$ will increase due to angular momentum transferring, the real disc accretion rate should decay faster than $1.2\pm0.2$. For most \{$M_\bullet$, $a_\bullet$\} pairs in the $1\sigma$ contour (see Figure 6 of W20), the fitted disc accretion rates decay as a power law with index $\sim 1.1$, but three cases have an index $<1$, $n=0.94\pm0.01$ for \{$8\times 10^6 M_\odot$, $0.998$\},  $n=0.90\pm0.01$ for \{$7\times 10^6 M_\odot$, $0.998$\}, and $n=0.95\pm0.02$ for \{$5\times 10^6 M_\odot$, $0.8$\}. These three pairs of \{$M_\bullet$, $a_\bullet$\}, located near the boundary of the $1\sigma$ contour, would fall outside the $1\sigma$ contour if we included the effect of a spreading disc, since the disc spreading effect will place an additional constraint on the $\dot m$ of each epoch. However, how much it improves the constraint on \{$M_\bullet$, $a_\bullet$\} is not the focus of this paper. 

As the power index ($\sim 1.1$) from the X-ray spectral fits is close to the index $1.2\pm0.2$ inferred from the observed optical/UV light curves (with no disc spreading effect), the viscous spreading of the disc should be slow. If not, the late time disc Optical/UV will be brighter than the observation due to a quick increasing disc size. We will explore this effect in more details in the following section.

\subsection{Constraints on Host Extinction and Disc Size}
\label{sec:host-ext}

\begin{table*}
\centering
\caption{Results of fits to the late time UV emission of ASASSN-14li.  
BH mass, BH spin, the disc accretion rate and inclination are fixed at the best-fit value obtained from the X-ray spectral fits (W20). The $\chi^2/d.o.f.$~values show that the use of the model {\sc xgalsb} (i.e., a starburst extinction curve) gives a significantly better fit than any another extinction model we consider, consistent with the post-starburst nature of the ASASSN-14li host galaxy. The final four rows of the table compare the effect of binning of the late-time UV data into different numbers of epochs.  We see that the addition of more groups make the fits slightly better, marginally warranting our inclusion of four groups. The fitted radius of the disc is about $2\times R_{\rm{t}}$ ($R_{\rm{t}}=10.13 R_{\rm{g}}$). Note that, lmc30dor \citep{Gordon+03}---$R_{\rm{V}} = 2.76$.
lmcavg \citep{Gordon+03}---$R_{\rm{V}} = 3.41$.
mwdense \citep{Cardelli+89}---$R_{\rm{V}} = 5.00$.
mwavg \citep{Cardelli+89}---$R_{\rm{V}} = 3.10$.
mwrv21 \citep{Cardelli+89}---$R_{\rm{V}} = 2.1$.
mwrv4 \citep{Cardelli+89}---$R_{\rm{V}} = 4.0$.
xgalsb \citep{Calzetti+00}---$R_{\rm{V}}=4.0$.}
\begin{tabular}{cccccccc}
\hline
Extinction model & $E_{\rm B-V}$ $[\rm{10^{-3} mag}]$ & $R_{1}$ $ [\rm{R_g}]$ & $R_{2}$ $ [\rm{R_g}]$  & $R_{3}$ $ [\rm{R_g}]$ & $R_{4}$ $ [\rm{R_g}]$ & $\chi^2/d.o.f$ \\
\hline
$\rm{NA}$
           &$\rm{NA}$   &$10.8 \pm 0.1$  &$\rm{NA}$ &$\rm{NA}$ &$\rm{NA}$  
           &$95.05/47$ \\
$\rm{lmc30dor}$
           &$0.2\pm0.1$   &$11.1\pm0.1$  &... &... &... 
           &$93.74/46$ \\
$\rm{lmcavg}$ 
           &$0.1\pm0.1$   &$11.1\pm0.1$  &... &... &...  
           &$94.43/46$ \\
$\rm{mwdense}$
          & $0.1\pm0.1$  &$11.0\pm0.1$  &... &... &...
           & $94.68/46$     \\
$\rm{mwavg}$ 
           & $0.1\pm0.1$  &$11.0\pm0.1$  &... &... &...
           & $94.61/46$     \\
$\rm{mwrv21}$
           &$0.1\pm0.1$   &$11.0\pm0.1$  &... &... &...   
           &$94.56/46$ \\
$\rm{mwrv4}$ 
           & $0.1\pm0.1$  &$11.0\pm0.1$  &... &... &...
           & $94.63/46$     \\
$\rm{xgalsb_1}$ 
           & $8.8\pm0.6$  &$19.2\pm1.0$  &... &... &...
          & $82.77/46$     \\
\hline
$\rm{xgalsb_2}$
           &$8.7\pm0.3$   &$19.2\pm1.0$  &$16.7\pm1.0$ &... &...   
           &$75.71/45$ \\
$\rm{xgalsb_3}$ 
           & $9.7\pm0.2$  &$20.0\pm2.0$  &$21.4\pm3.0$  &$17.8\pm2.0$ &...
           & $69.73/44$     \\
$\rm{xgalsb_4}$ 
           &$9.8\pm0.2$ &$20.1\pm4.1$  &$21.4\pm3.0$  &$21.9\pm4.1$ &$17.8\pm3.0$
           & $69.20/43$     \\
\hline
\end{tabular}
\label{Tab:uv}
\end{table*}

In this section, we will explore whether the host extinction model affects the UV fitting results. Furthermore, we investigate how the other free parameter, the disc radius $R_{\rm out}$, evolves in time. For the first question, we fit the four UV light curves with the slim disc model attenuated by different extinction models. For the second question, we fit the light curves employing different data groups and different values of the $\{ M_\bullet$, $a_\bullet\}$ pair using the slim disc model (including our fiducial extinction model).

Table~\ref{Tab:uv} shows the fitting results for different extinction models and different data groupings. We first fit the data with the same $R_{\rm out}$ for all epochs and no extinction (row 1). The fits to the data are poor, as the reduced $\chi^2$ values are $>2$ and the fitted disc outer radius is $\approx 10.8 R_{\rm g}$, which is $\approx 1.1 R_{\rm t}$\footnote{For $M_\bullet=10^7 M_\odot$, $R_{\rm t}$ is about $R_{\rm t} \approx 10.13 R_{\rm g}$ for a disruption of a solar-like star with penetration parameter $\beta=1$.}. This 
is in tension with both the theoretical expectation that $R_{\rm out}=2R_{\rm t}/\beta$
(although it could be explained by a deeply plunging TDE), and also the X-ray spectral fits. As most X-ray photons are generated within $20 R_g$ (W20), a much smaller disc size would affect the X-ray spectrum significantly. The (too) small fitted disc size from a model that ignores the effects of extinction on the emergent UV spectrum indicates that host extinction is likely important. 

In order to fit the host extinction, we fit the light curves with 7 different extinction curves from {\sc pysynphot} \citep{Lim+15}; see row 2-8 of Table~\ref{Tab:uv}. The impact on the UV emission can be described with only one parameter ($E_{\rm B-V}$) for all the 7 extinction models. From Table~\ref{Tab:uv}, one can see that the LMC and MW extinction models do not yield a better fit to the data when compared with no model for the extinction, as the $\chi^2$ only decrease by $\approx 1$ which is not significant given that there is one additional free parameter. Furthermore, the best-fit $E_{\rm B-V}$ values are consistent with 0 at the $1 \sigma$ confidence level, and the fitted $R_{\rm out}$ is marginally consistent with the value obtained without modeling for the effect of extinction. However, the starburst extinction model results in a better fit with $\chi^2$ deceasing by $\sim 12$. The disc size has a best-fit value of $19.2\pm 1.0 R_{\rm g}$, which is consistent with $R_{\rm out} = 2R_{\rm t}$ ($20.3 R_{\rm g}$). The fitted $E(B-V)$ is about 0.009 mag. This low value is not a surprise since: (1) surveys show that the value of $E(B-V)$ is within the range of $0.0$--$1.0$ mag often found in starburst galaxies \citep{Calzetti1997}, (2) the low $E(B-V)$ value is consistent with the UV detection of ASASSN~14li, (3) we know that ASASSN-14li occurred in a post-starburst (\citep{Prieto2016}) galaxy that ended $\sim$400 Myr ago \citep{French2017}
and that the dust content of such galaxies declines over time \citep{Li2019}.

\begin{figure}
\includegraphics[height=.28\textheight]{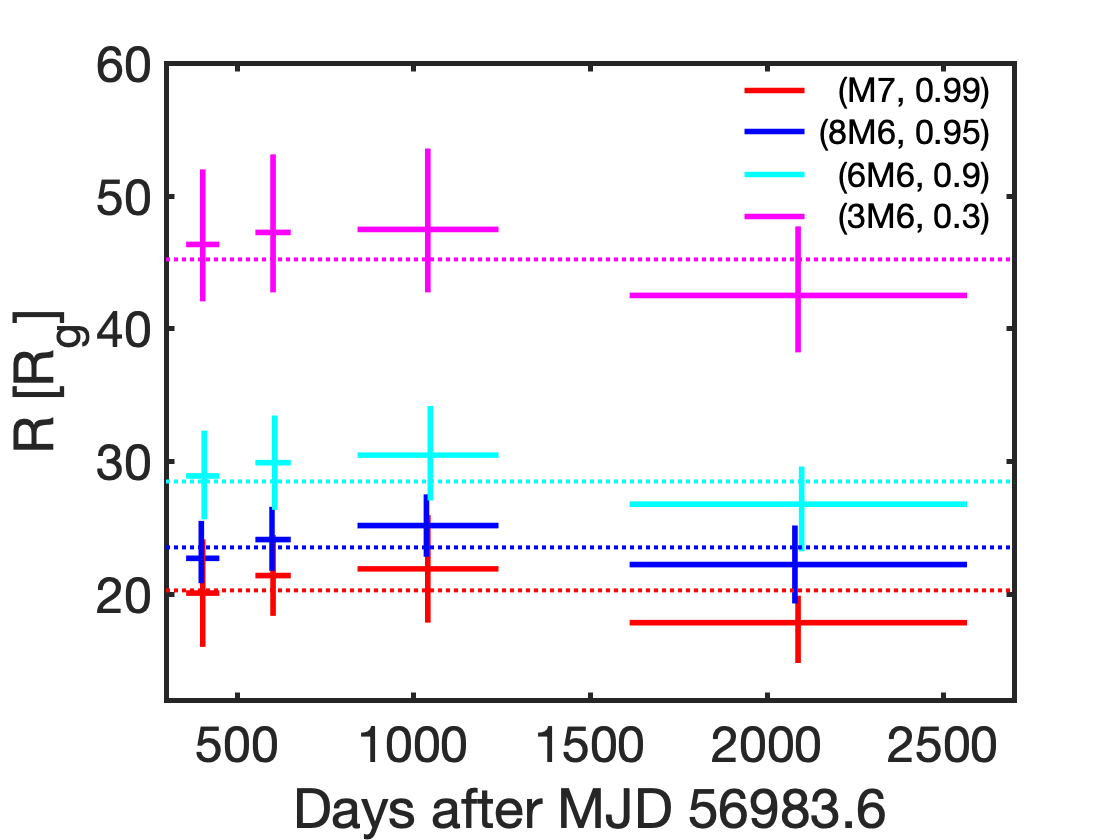}
\caption{Evolution of $R_{\rm out}$ expressed in unit of $R_g$ for different pairs of $\{ M_\bullet, a_\bullet\}$. We fit the UV light curves with a slim disc plus starburst extinction model. The accretion rate and disc inclination are fixed at their best-fit value from X-ray spectral fitting. The dotted colour lines show $R_{\rm out}=2R_{\rm t}$, respectively. We find that $R_{\rm out}$ is consistent with a constant value at the $1\sigma$ confidence level, for all four cases, challenging the baseline expectation of a viscously spreading disc. 
}
\label{Fig:Rout}
\end{figure}

We now explore the evolution of disc size. In rows ${\rm xgalsb_2}$, ${\rm xgalsb_3}$ and ${\rm xgalsb_4}$ of Table \ref{Tab:uv}, we fit the light curves with 2 groups, 3 groups, and 4 groups, respectively. When grouping the data, we always set the last two epochs as a separate group, as the accretion rates of these two epochs cannot be determined independently (there are no X-ray observations at those epochs). 
For ${\rm xgalsb_2}$, the $\chi^2$ decreases by about 7 when compared to the best-fit value for the ${\rm xgalsb_1}$ model. The $p$ values obtained from an F-test \citep{Conder2020} relative to the 1-group model are 0.05, 0.03, and 0.05, for 2-, 3-, and 4-group models, respectively.\footnote{p is the probability (between 0 and 1) that the improvement of the fit is due to chance. Therefore, a small value of p means a high confidence that the additional parameters are warranted.} These $p$ values show that the additional groups are only marginally warranted. The $\Delta {\rm AIC}$ \citep{Akaike1974} relative to a 1-group model are 5.0, 9.0, and 7.5, for 2-, 3-, and 4-group models, respectively \footnote{
$\Delta{\rm AIC}=\Delta \chi^2+2\Delta K$,  
where $K$ is the number of free parameters. Generally speaking, two models with $\Delta {\rm AIC}=5$ and $10$ are considered to present strong and very strong evidence, respectively, against the weaker model.}.  These $\Delta {\rm AIC}$ values also show that adding more data bins is not strongly preferred. These two tests show that there is no strong evidence for evolution of $R_{\rm out}$ at different times. This can be also seen from the fitted radii of the 4-group model (see also Fig.~\ref{Fig:Rout}). In each of the 4 temporal groups, the outer radius is consistent with $2R_{\rm t}$ at a $1\sigma$ level.

Figure \ref{Fig:Rout} plots the fitted $R_{\rm out}$ for different pairs of $M_\bullet$ and $a_\bullet$. As one can see the fitted $R_{\rm out}$ is consistent with a constant value at the $1\sigma$ confidence level for all cases considered. In each case, the fitted $R_{\rm out}$ is consistent with $2R_{\rm t}$.
As the debris is accreted, it will transfer part of its angular momentum to the outer disc, which (absent additional physics) will result in viscous expansion.

However, our fits find significantly less viscous spreading than would be expected in simple models.  This could signify that either (1) the majority of the bound debris remains unaccreted in a low-viscosity disc at late epochs, or (2) significant amounts of disc angular momentum have been expelled from the system during the circularization\footnote{See e.g. \citep{BonnerotLu19} for an interesting example of how circularization may preferentially expel matter in one direction, dramatically changing the angular momentum budget of the remaining bound material.} or accretion process.  The latter possibility may be the most natural; angular momentum loss in a magnetized wind has been predicted in both analytic \citep{BlandfordPayne82, FerreiraPelletier95} and numerical \citep{Scepi+18} magnetohydrodynamics.

In general, our fits show that, 1) only a starburst extinction curve can give a better and more reasonable fit to the observed optical/UV light curves, 2) the disc size is about 2$R_{\rm t}$, and is consistent with a constant at the $1\sigma$ CL.

\section{Implications of ASASSN-14\lowercase{li's}
Early-Time Optical/UV Light Curves}
\label{earlylc}

\begin{figure}
\includegraphics[height=.28\textheight]{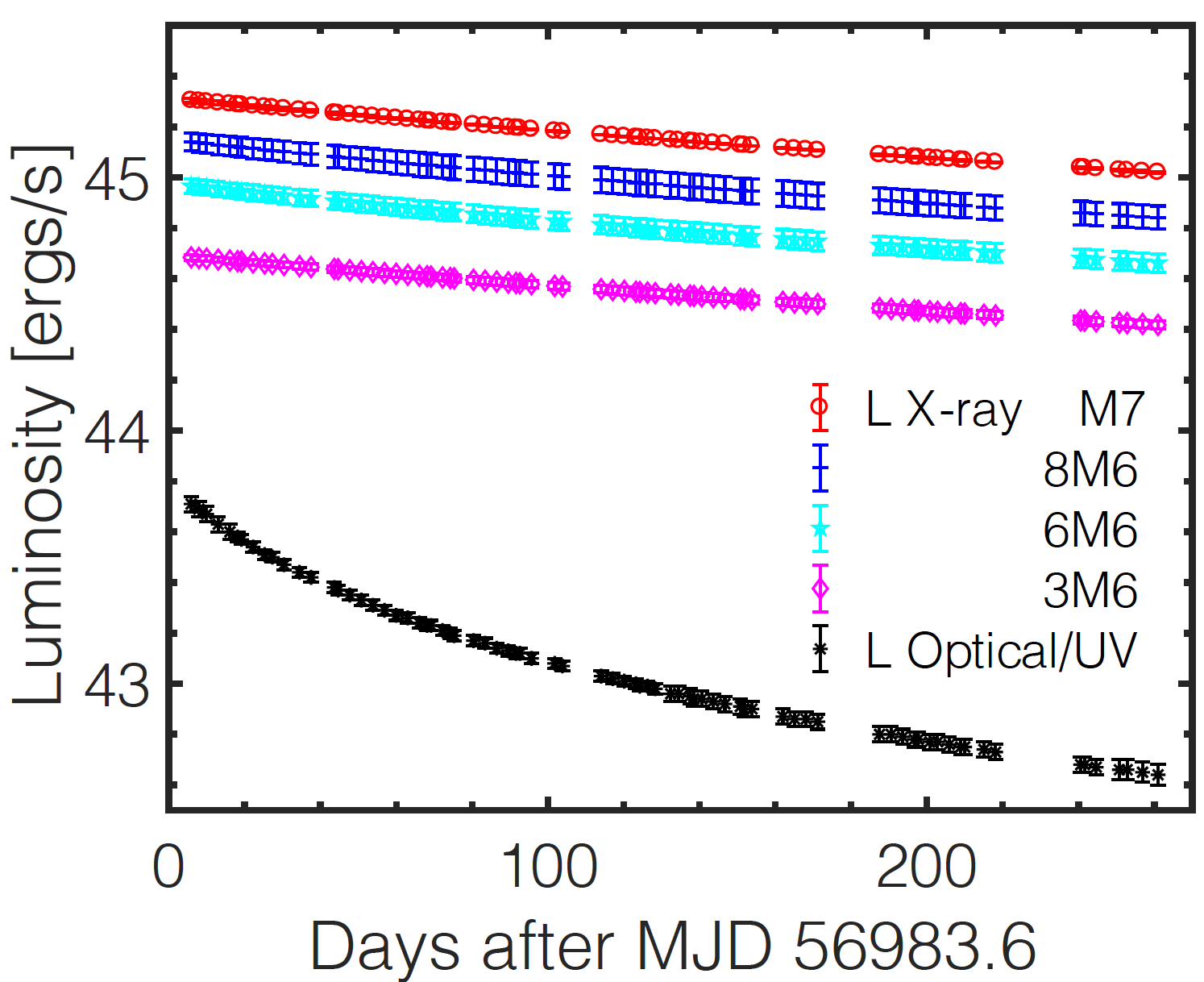}
\caption{The early time bolometric luminosity derived from slim disc modeling and the blackbody Optical/UV emission. For the bolometric luminosity inferred from X-ray spectral fitting, we estimate $\dot m$ at each Optical/UV epoch by using the decay of accretion rate from X-ray spectra fitting. } 
\label{Fig:Luvx}
\end{figure}

As is shown in Figure~\ref{Fig:lateUV}, the early time UV emission is at least 10 times brighter than the predicted near-UV emission from our slim disc fits (which peak at $\sim 0.1$ keV) to early time X-rays. On the other hand, the bolometric early time disc luminosity is $>10$ times brighter than the luminosity derived from the early time blackbody near-UV emission (see Fig.\ref{Fig:Luvx}). These two factors indicate that there should be another thermal photosphere, with a lower effective temperature ($\sim 3\times 10^4$ K) but much larger emitting area than the disc, in order to account for the early UV emission. The origin of this large-scale, early time photosphere (which is typical of optically selected TDEs) has been the subject of much theoretical debate; see \citep{Roth+20} for a recent review.

The rich X-ray dataset available for ASASSN-14li allows us to test competing models for early-time optical/NUV emission in greater detail than is possible for most other TDEs observed to date.  In this Section, we present a set of self-consistency checks that can be employed to check the validity of (i) shock-powered early-time emission \citep{Piran+15, Shiokawa+15, SteinbergStone22}; (ii) reprocessing of X-ray/EUV photons by an outflowing wind \citep{Metzger2016, RothKasen18, LuBonnerot19, Piro+20}; and (iii) reprocessing of X-ray/EUV photons by a quasi-static photosphere \citep{Loeb+97, Coughlin+2014, Guillochon+14, Roth+16}. The test of the shock-powered emission mechanism is entirely novel, while the tests of the two variants of the reprocessing paradigm follow a similar line of reasoning as in the work of \citep{MatsumotoPiran20}.

\subsection{Emission from a Self-Intersecting Shock}
\label{sec:shocks}
In W20, we first presented detailed, time-dependent fits to the XMM-{\it Newton} spectra of ASASSN-14li, which have been slightly updated here.  We can now combine this time-dependent estimate for the ionizing continuum produced by the inner disc with independent optical/NUV light curves \citep{Holoien2016a, Brown+17} to pose a simple geometrical question.  The optical/NUV light curves of ASASSN-14li imply an emitting area which can be measured directly under the assumption of a blackbody spectrum.  In the simplest version of the shock paradigm, the radial location of the optical/NUV photosphere is linked to the physical position of the stream self-intersection shocks, and therefore we may determine the total solid angle subtended by the optical/NUV photosphere (from the perspective of the central accretion disc).  This occluded solid angle, in combination with the ionizing continuum of the central disc, determines a characteristic reprocessing luminosity which must be significantly less than the observed optical/NUV luminosity in order for shock-powered emission to provide a self-consistent model.

To make the above self-consistency check more quantitative, we will consider a star of mass $M_\star$ and radius $R_\star$ disrupted by a SMBH of mass $M_\bullet$ (and spin $a_\bullet$).  Tidal disruption occurs if the star's pericenter $R_{\rm p}$ is less than the tidal radius \citep{Rees1988} 
\begin{equation}
    R_{\rm t} = R_\star \left( \frac{M_\bullet}{M_\star}\right)^{1/3}.
\end{equation}
Equivalently, disruption occurs if $\beta = R_{\rm t}/R_{\rm p} > 1$.  The elongated debris streams produced following disruption fly out to a large range of radii before returning to pericenter \citep{Stone+13, GuillochonRamirezRuiz13}; the semimajor axis of the most tightly bound debris is
\begin{equation}
    a_{\rm min} = \frac{R_{\rm t}^2}{2R_\star} = \frac{R_\star}{2} \left( \frac{M_\bullet}{M_\star} \right)^{2/3}.
\end{equation}
Shocks will be generated by the self-intersection of debris streams\footnote{In the recent simulations of \citep{SteinbergStone22}, shock-powered light curves are driven by dissipation at radii much smaller than $R_{\rm SI}$, but as we shall see, such a situation would only {\it strengthen} the constraints derived in this sub-section.}, at a radius $R_{\rm SI}$.  If the unfortunate star had a highly relativistic pericenter, apsidal precession will force self-intersections at $R_{\rm SI} \ll a_{\rm min}$, while a less relativistic pericenter will result in quasi-apocentric self-intersections\footnote{Note that this picture may be complicated if relativistic nodal precession is sufficient to prevent immediate stream self-intersections \citep{GuillochonRamirezRuiz15, Hayasaki+16}.  Even in this case, however, most of the bound debris would orbit on trajectories with $a \sim a_{\rm min}$.} at $R_{\rm SI} \approx a_{\rm min}$ \citep{Dai+15}.  More precisely, the debris stream longitude of pericenter $\omega$ will, at leading post-Newtonian order, shift by an amount \citep{Merritt+10}
\begin{equation}
    \delta \omega = A_{\rm S} - 2A_{\rm J}\cos \iota
\end{equation}
per orbit (with most of the apsidal shift occurring during pericenter passage).  Here $\iota$ is the spin-orbit misalignment angle, and the individual terms in this equation are
\begin{align}
    A_{\rm S} =& \frac{6\pi}{c^2} \frac{GM_\bullet}{R_{\rm p}(1+e)} \approx 11.5^\circ \left(\frac{\tilde{R}_{\rm p}}{47.1} \right)^{-1}     \\
    A_{\rm J} =&\frac{4\pi}{c^3} a_\bullet \left(\frac{GM_\bullet}{R_{\rm p}(1+e)}\right)^{3/2}    \approx 0.788^\circ \left(\frac{\tilde{R}_{\rm p}}{47.1} \right)^{-3/2}    a_\bullet,
\end{align}
representing the leading-order contributions of the SMBH mass and spin, respectively.  Note that $\tilde{R}_{\rm p} = R_{\rm p} / R_{\rm g}$ is a dimensionless pericenter normalized by the gravitational radius ($R_{\rm g} = GM_\bullet / c^2$), and  $e$ is the eccentricity of the debris streams (typically, $0.99 < e <1$).  From the above equations we see that the contribution of SMBH spin $a_\bullet$ to $\delta \omega$ is always highly sub-dominant for $\tilde{R}_{\rm p} \gtrsim 10$ and therefore we neglect it in the remainder of this calculation.  The apsidal shift $\delta \omega$ causes stream self-intersection at the radius \citep{Dai+15}
\begin{equation}
    R_{\rm SI} = \frac{R_{\rm p}(1+e)}{1+e\cos(\pi + \delta \omega /2)}.
\end{equation}
Under the assumption that the photosphere is located near the location of the shocks, the total area of the celestial sphere at the photospheric radius will be $A_{\rm SI} = 4\pi R_{\rm SI}^2$.

Under the standard assumption that observed optical/UV emission is a thermal blackbody, we may combine multiband photometric observations to estimate the {\it true} area of the optical photosphere, $A_{\rm bb} = L_{\rm bb} / (\sigma_{\rm SB} T_{\rm bb}^4)$.  Here $\sigma_{\rm SB}$ is the Stefan-Boltzmann constant, and $L_{\rm bb}$ and $T_{\rm bb}$ are the observationally inferred blackbody luminosity (bolometric) and temperature.  We now compute the covering fraction
\begin{equation}
    f_{\rm A}(t) = \frac{A_{\rm bb}(t)}{A_{\rm SI}}=\frac{A_{\rm bb}(t)}{4\pi R_{\rm SI}^2},
    \label{fa}
\end{equation}
which in turn can be used to relate the central ionizing luminosity $L_{\rm disc}$ to the time-dependent reprocessing luminosity
\begin{equation}
    L_{\rm rep}(t) = f_{\rm A}(t) L_{\rm disc}(t).
    \label{lr}
\end{equation}
We now have two related (but distinct), non-trivial, multi-epoch self-consistency checks for the shock paradigm.  Shock-powered emission will not be self-consistent if $f_{\rm A} \gg 1$, nor if $L_{\rm rep}(t) > L_{\rm bb}(t)$.  The former inconsistency would imply that the observed optical/NUV photosphere is much larger than plausible self-intersection radii (and as a result, plausible shock dissipation sites will not be able to produce the observed black body radiation).  In principle one could accommodate this by positing that the post-shock material must expand significantly before trapped photons can escape, but in practice this will reduce the theoretical shock-powered luminosity below what is observed (due to adiabatic degradation).  The latter inconsistency would imply that the observed ionizing continuum, in combination with the observed optical/NUV photosphere, will naturally produce the observed optical/NUV luminosity through reprocessing alone.  

These consistency checks depend on both time-dependent quantities, such as $L_{\rm disc}(t)$, $L_{\rm bb}(t)$, and $T_{\rm bb}(t)$ and also inferred (or assumed) parameters of the TDE: $M_{\bullet}$, $\beta$, $M_\star$, and $R_\star$.  We use our X-ray spectral modeling from W20 to estimate ranges of $M_\bullet$, and the associated $L_{\rm disc}(t)$ curves (which will change as $M_\bullet$ changes). 
$R_\star$ is determined from $M_\star$ using main sequence mass-radius relationships \citep{KippenhahnWeigert90}, and $M_\star$ is left as a free parameter.  We note that UV emission line ratios provide significant constraints on the mass of the ASASSN-14li progenitor star due to the impact of CNO-cycle burning \citep{Kochanek16, Yang+17}, and this reasoning has been used to argue that the progenitor of ASASSN-14li was at least $\approx 1.3 M_\odot$ \citep{Mockler+22}. 
However, from the decay rate of $\dot m$ (see Fig.\ref{Fig:mdot}), it is difficult to determine whether this star is fully or partially disrupted. In addition, from the late time Optical/UV light curves fitting, we get $R_{\rm out}\sim 2R_t$ (see Fig.\ref{Fig:Rout}). These two factors indicate that $\beta\sim1$ is highly possible. As a result, we take $\beta=1$ ($R_{\rm p}=R_{\rm t}$) as an assumption (if $\beta$ were still larger, the consistency checks would become harder to satisfy).

\begin{figure*}
\centering
\includegraphics[height=.55\textheight]{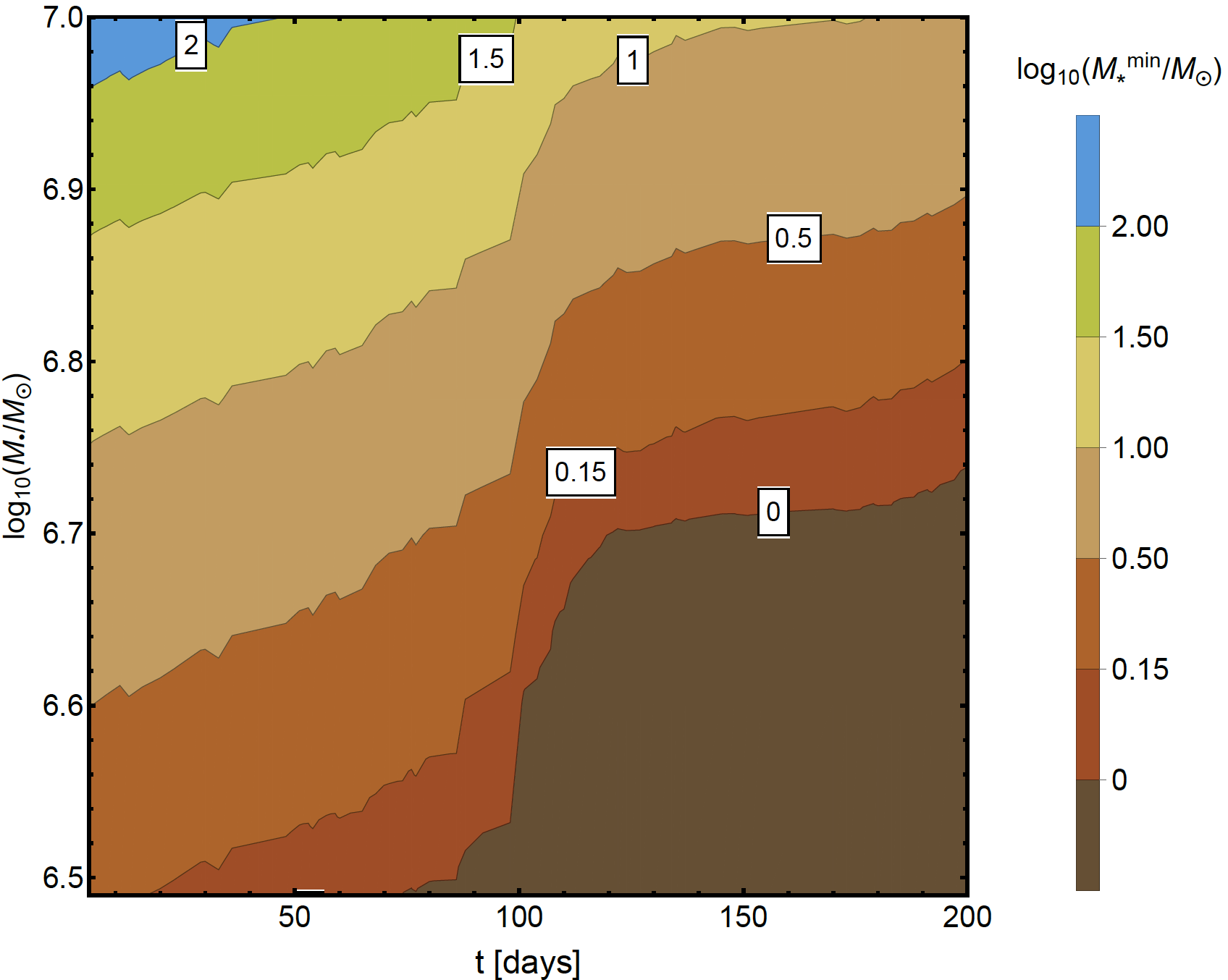}
\caption{Contours indicating the {\it minimum} mass of the disrupted star, $M_\star$, required for the shock paradigm to provide a self-consistent description of early-time optical emission.  Contours are colour-coded and labeled in terms of $\log_{10}(M_\star / M_\odot)$.  Different minimum masses are required at different observing epochs (shown on the x-axis as time since first observations), and the minimum mass also depends on the SMBH mass $M_\bullet$, which is constrained by X-ray continuum fitting to the range shown on the y-axis (at a 1$\sigma$ confidence level).  The minimum mass of the disrupted star required for self-consistency of the shock paradigm at all times is $M_\star \approx 10^{0.15}M_\odot \approx 1.4 M_\odot$.  Unrealistically massive stars are needed if $M_\bullet \gtrsim 10^{6.7}M_\odot$.
}
\label{fig:shockConstraints}
\end{figure*}

\begin{figure}
\includegraphics[height=.22\textheight]{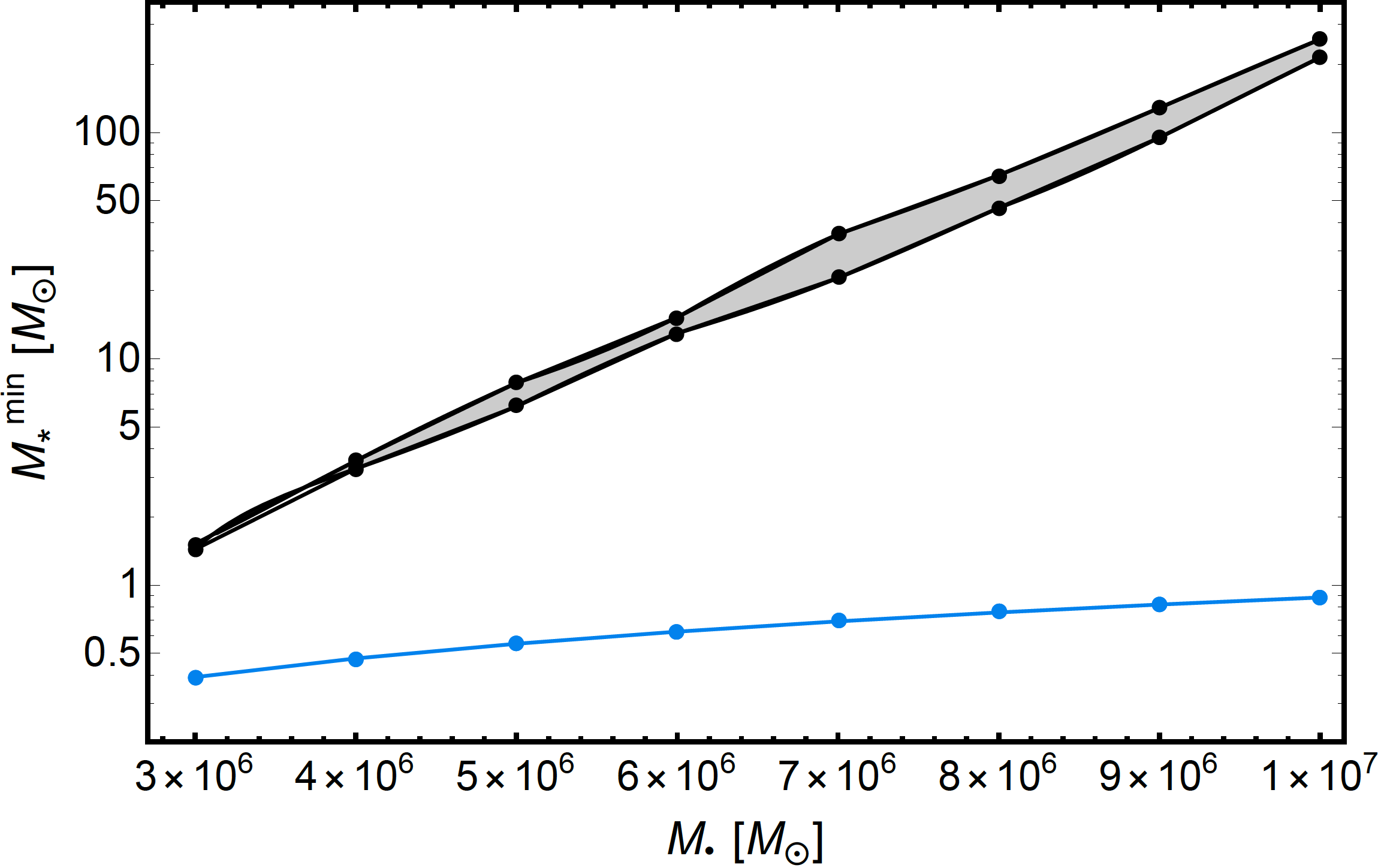}
\caption{Minimum stellar mass, $M_\star^{\rm min}$, required for the shock paradigm to be self-consistent.  The self-consistency requirement that $L_{\rm rep} \le L_{\rm bb}$ at all observing epochs is shown in black (shaded error region corresponds to the range of permitted SMBH spins at each mass found via X-ray fitting), and the weaker requirement that the photosphere covering fraction $f_{\rm A} \le 1$ at all epochs is shown in blue.  Both constraints are shown at their strongest, i.e. in the earliest epochs.  Reasonable stellar masses (i.e. $\lesssim 3 M_\odot$, as should be typical for a post-starburst host) allow self-consistency when the SMBH mass is smaller, but a large portion of the 1$\sigma$ CL on $M_\bullet$ (obtained from disc continuum fitting) can be ruled out if the shock paradigm is the true origin of optical/NUV luminosity in ASASSN-14li.
}
\label{fig:simpleShockConstraints}
\end{figure}

To test the self-consistency of the shock paradigm, we require $f_{\rm A} \le 1$ (Eq.~\ref{fa}) and $L_{\rm rep} \le L_{\rm bb}$ (Eq.~\ref{lr}). We use these upper limits on $f_{\rm A}$ and $L_{\rm rep}$ to place lower limits on the unknown mass of the victim star, $M_\star$. 
For ASASSN-14li\footnote{For a different TDE with a lower disc luminosity, the $f_{\rm A}\le 1$ consistency check could be the more constraining of the two.}, the second self-consistency test ($L_{\rm rep} \le L_{\rm bb}$) is at all epochs more constraining than the first ($f_{\rm A} \le 1$), due to the fact that $L_{\rm disc} \gg L_{\rm bb}$. 
We show the resulting lower limits on $M_\star$ in Fig. \ref{fig:shockConstraints}.  As the absolute strongest constraints on the shock paradigm come from the earliest epochs, we present these alone in Fig. \ref{fig:simpleShockConstraints}.

We see that the self-consistency of the shock paradigm is most constrained by early-time observations ($t \lesssim 100$ days after detection).  The resulting lower limits are a sensitive function of SMBH mass.  The most generous lower limits are achieved for the smallest SMBHs; at the bottom end of the 1$\sigma$ confidence limit on $M_\bullet$ ($10^{6.5} M_\odot$), the shock paradigm is self-consistent provided so long as $M_\star \gtrsim 1.4 M_\odot$, which is plausible.  Conversely, in the upper half of the $1\sigma$ CL SMBH mass region ($M_\bullet  \gtrsim 10^{6.75}M_\odot$), the shock paradigm requires $M_\star \gtrsim 10 M_\odot$, necessitating a very rare TDE which is unlikely to occur in the small present-day TDE sample.

\subsection{Emission from a Reprocessing Layer}

In this section, we will test whether the early optical/UV emission can be produced by a reprocessing layer of gas.
As the disc luminosity is $>10$ times brighter than the optical/UV black body luminosity, the putative reprocessing layer is either optically thin (reprocessing only a small fraction of the photons), or optical thick but only covering a small fraction of the emitted sphere. However, as observations of TDE optical/UV emission are generally consistent with optically thick thermal emission \citep{Dai+18,Roth+16}, we only consider the latter possibility, and follow the work of \cite{MatsumotoPiran20} to estimate how much mass is required to support such a reprocessing layer. As before, we assume an optically thick reprocessing layer with a covering fraction 
\begin{equation}
    f_{\rm A} (t)=\frac{L_{\rm bb}(t)}{L_{\rm disc}(t)}.
    \label{coverfactor}
\end{equation}
Such a gas layer can be produced by outflowing material, either from a super-Eddington disc wind or from the result of shocks in stream-stream or stream-disc collisions \citep{Metzger2016, RothKasen18, Dai+18, LuBonnerot19}.  Alternatively, it can be formed from bound but poorly circularized debris \citep{Loeb+97, Guillochon+14, Roth+16}. Here we estimate the mass required for both outflowing and quasi-static reprocessing layer models in ASASSN-14li.  Our aim is to determine whether reprocessing models can operate within the mass budget provided by the disruption of a plausible star (i.e. $\lesssim 1 M_\odot$).

\subsubsection{Outflowing Debris Layer}

We follow the work of \cite{MatsumotoPiran20}  to estimate the mass of the outflowing material with different speeds. We note that similar models \citep{Metzger2016,RothKasen18} have been developed for cases of high speed outflowing material ($\sim10,000$km/s). 
We introduce two critical radii to understand the emergent spectrum, as in \citep{Shen+15}. The first is the thermalization radius, $r_{\rm th}$, below which the radiation can reach local thermal equilibrium with the gas. We note that the thermalization radius is larger than the radius corresponding to the observed luminosity and black body temperature ($r_{\rm bb}=\sqrt{L_{bb}/(4\pi\sigma_{SB}T^4_{bb})}$), due to the fact that the electron scattering optical depth 
$\tau_{\rm es} > 1$ \citep{Roth+16}.  At this radius, the effective absorption depth, $\tau_{\rm eff}$, is equal to unity \citep{Rybicki+79}\footnote{While $\tau_{\rm eff}$ is properly defined as a frequency-dependent variable, the photons around the Wien peak dominate the luminosity \citep{Rybicki+79}. We follow the work of \citep{Shen+15} and define $\tau_{\rm eff}$ to be the effective optical depth of the photons near the Wien peak.}: 
\begin{equation}
\tau_{{\rm eff}}(r_{\rm th})=\int_{r_{\rm{th}}}^{ r_{\rm{o}}}\rho(r)\sqrt{3\kappa_{\rm a}(\kappa_{\rm a}+\kappa_{\rm es})}\,dr=1,
\label{taueff}
\end{equation}
here, $\kappa_{\rm es} \approx 0.35 ~{\rm cm}^{2} {\rm g}^{-1}$ is the Thomson electron scattering opacity, $r_{\rm o}$ is the outer edge of the layer, and $\kappa_{\rm{a}}$ is the absorption opacity. In order to derive $r_{\rm th}$, we use Kramer's law to approximate the absorption opacity,
\begin{equation}
\kappa_{\rm{a}}=\kappa_0\frac{\rho(r)}{\rm g/cm^3}\left(\frac{T}{K}\right)^{-7/2}~\rm{cm^2~g^{-1}},
\label{ka}
\end{equation}
where $\kappa_0$ is a constant which also contains the dependence on the
gas composition. Following \cite{MatsumotoPiran20}, we set $\kappa_0=4.0\times 10^{25}$ for TDEs, a normalization which is consistent with more precise calculations of the Planck mean absorption opacity at Solar metallicity.

The second critical radius is the photon trapping radius $r_{\rm tr}$, above which the photon diffusion time is shorter than the dynamical time and photons can freely escape from the outflowing material. This radius is defined as,
\begin{equation}
\tau(r_{\rm tr})=\int_{r_{\rm{tr}}}^{r_{\rm{o}}}(\kappa_{\rm a}+\kappa_{\rm es})\rho\,dr=\frac{c}{v_w},
\label{tau_es1}
\end{equation}
where c is the speed of light and $v_{\rm w}$ is the wind speed (at $r_{\rm tr}$). When $r_{\rm tr} > r_{\rm th}$, the photons are trapped with the outflowing material until reaching $r_{\rm tr}$. As a result, the observed luminosity and observed black body temperature are given by the diffusion luminosity and gas temperature at $r_{\rm tr}$, respectively.

We now model the density $\rho(r)$ at $r_{\rm tr}$ and $r_{\rm th}$. As in \citep{MatsumotoPiran20}, we consider a steady wind outflow of constant speed $v_{\rm w}$, but unlike most earlier analytic work, we assume a covering factor $f_{\rm A}$ that may be less than unity.  More specifically, we assume that the outflow 
fills a solid angle on the sky that is $\Delta \Omega = 4\pi f_{\rm A}$, and therefore the density profile along a ray is 
\begin{equation}
\rho(r) = \frac{\dot{M}_{\rm wind}}{4\pi r^2 v_{\rm w} f_{\rm A}}.    
\end{equation}
As $\kappa_{\rm a}\ll \kappa_{\rm es}$ for TDE photospheres \citep{Roth+16} and the density decays faster than $r^{-1}$ 
in a general outflow, we approximate the radial optical depth in Eq.~\ref{taueff} and Eq.~\ref{tau_es1} as,
\begin{align}
\label{tau_eff}
&\tau_{\rm eff}(r_{\rm th})\approx\sqrt{3\kappa_{\rm a}\kappa_{\rm es}}\rho_{\rm th} r_{\rm th}=1, \\
&\tau(r_{\rm tr})\approx\kappa_{\rm es}\rho_{\rm tr} r_{\rm tr}=\frac{c}{v_w}.
\label{tau_es2}
\end{align}

Assuming thermal equilibrium between photons and the gas, and applying the diffusion approximation, the gas temperature at a given radius in a spherically symmetric outflow can be described as,
\begin{equation}
\frac{{\rm d}(T^4)}{{\rm d} r}=-\frac{3\kappa_{\rm es}\rho(r)}{16 \pi \sigma_{\rm SB}r^2} L_{\rm disc}(t).
\label{dt}
\end{equation}
We use $L_{\rm disc}(t)$ instead of $L_{\rm bb}(t)$ because the covering factor $f_{\rm A}$ that we consider may be less than unity.  We note that the prior equation is only strictly valid in the limit of spherical symmetry, and may break down due to anisotropic photon diffusion in non-spherical systems with $f_{\rm A} \ll 1$ (in this case Eq. \ref{dt} will overestimate the temperature at a given radius).

Taking the approximation ${\rm d} (T^4)/{\rm d} r\approx- T_{\rm bb}(t)^4/r$ at $r=r_{\rm tr}$ (or at $r=r_{\rm th}$) and using Eqs. \ref{tau_eff} and \ref{tau_es2}, we get the outflow rate 
\begin{equation}
\begin{split}
\dot M_{\rm wind}=\left \{
\begin{array}{ll}
f_{\rm A}(t)^{\frac{1}{5}}L_{\rm bb}(t)^{\frac{4}{5}} T_{\rm bb}(t)^{-\frac{11}{10}} v_w \left(\frac{3\pi \kappa_{\rm es}}{2^{6}\sigma_{\rm SB}^4 \kappa_0^3}\right)^{\frac{1}{5}},     & \rm{if} ~v < v_{\rm c},\\
f_{\rm A}(t)^{\frac{1}{2}}L_{\rm bb}(t)^{\frac{1}{2}} T_{\rm bb}(t)^{-2} v_w^{-\frac{1}{2}}\left(\frac{3\pi c^3}{\sigma_{\rm SB} \kappa_{\rm es}^2}\right)^{\frac{1}{2}},      & \rm{if} ~v > v_{\rm c}.
\label{mlb}
\end{array}
\right.
\end{split}
\end{equation}
Here, the critical velocity, $v_{\rm c}$, as determined by $r_{\rm th}=r_{\rm tr}$, can be estimated as
\begin{equation}
v_c=f_{\rm A}(t)^{\frac{1}{5}}\left( \frac{48\pi \sigma_{\rm SB} \kappa_0^2 c^5}{\kappa_{\rm es}^4}\right)^{\frac{1}{5}} L_{\rm bb}(t)^{-\frac{1}{5}} T_{\rm bb}(t)^{-\frac{3}{5}}.    \label{vc}
\end{equation}
We note that the $\dot M$ and $v_c$ derived here are different from those in \cite{MatsumotoPiran20} by a factor of $f_{\rm A}(t)^{1/5}$ (or $f_{\rm A}(t)^{1/2}$), primarily due to the covering factor\footnote{There is also a factor of 3 in the definition of effective optical depth.}. 

The mass lost by outflowing material, from the discovery day ($t_{\rm d}$) to the current epoch $t$, can be calculated as,
\begin{equation}
M_{\rm{wind}}=\int_{t_{\rm{d}}}^{t}\dot M_{\rm wind}\,dt'.
\label{wind}
\end{equation}

\begin{figure}
\centering
\includegraphics[height=0.28\textheight]{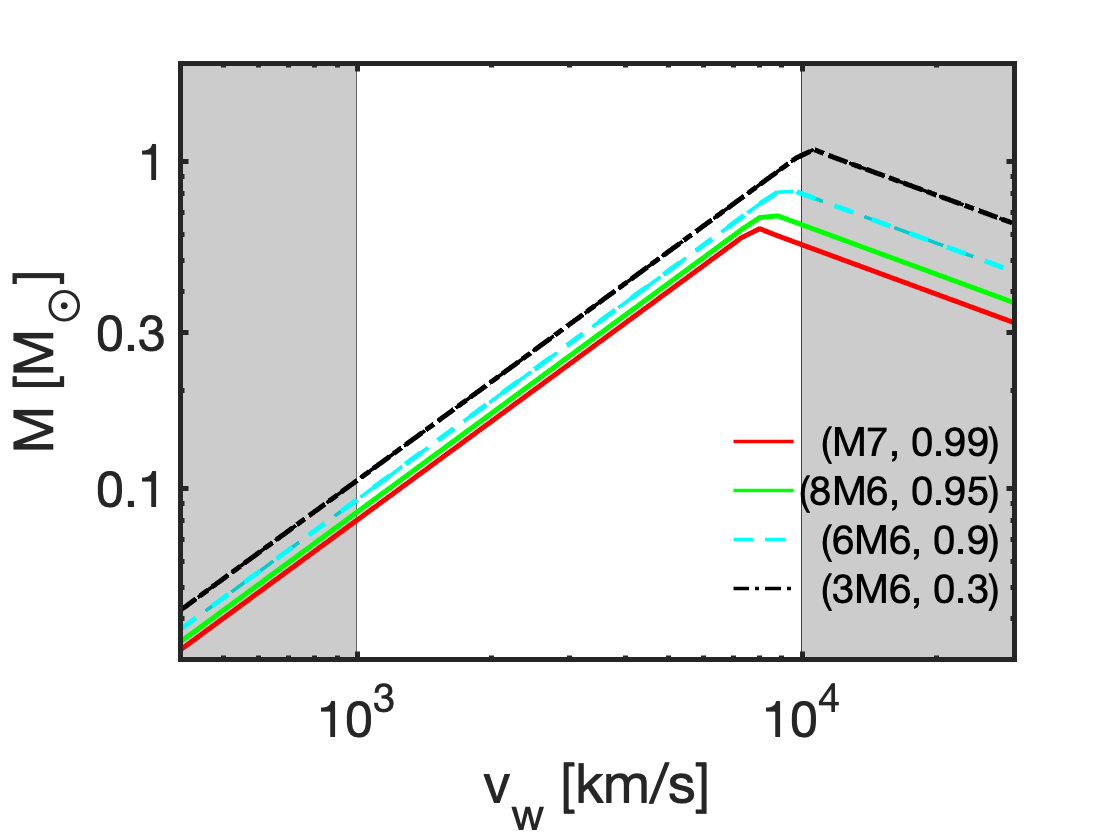}
\caption{Mass lost by outflowing material for different velocities. The gray shaded regions show velocities excluded by the observations
of line widths $v_w<10^4$  $\rm{km/s}$ \citep{Arcavi+14} or lower than the escape velocity $v_w \sim 10^3$  $\rm{km/s}$. Even though the full width at half maximum (FWHM) of the line is likely determined by effects of electron scattering, the velocity determined by the FWHM can not be lower than the outflow velocity. The peak of each line denotes $v_{\rm w}=v_{\rm c}$.}
\label{Fig:wind}
\end{figure}

Figure \ref{Fig:wind} shows the mass lost by outflow for different wind velocities varying from $4\times 10^2$ km/s to $4\times 10^4$ km/s. The estimated amount of mass lost peaks at the critical velocity of $v_{\rm w}\sim 10^4$ km/s for which $r_{\rm th}=r_{\rm tr}$ and it decreases with increasing $M_\bullet$. This is because a higher $M_\bullet$ yields a larger disc, which in turn implies a brighter $L_{\rm disc}$, which makes the covering factor smaller. The maximum of the amount of mass lost of each curve varies between 0.6 and 1.1 $M_\odot$, which is somewhat smaller than the $\sim$few $M_\odot$ derived by \cite{MatsumotoPiran20}. This difference is mainly caused by the covering factor. In this work, we found that the disc luminosity is $>10$ times brighter than the observed optical/UV black body luminosity at peak, which indicates $f_{\rm A}(t_p)<0.1$, while \cite{MatsumotoPiran20} assume $f_{\rm A}(t)=1$ ($L_{\rm disc}=L_{\rm bb}$). 
As the amount of mass lost is between $\sim 0.1 M_\odot$ and $\sim 1.1 M_\odot$, we cannot rule the outflow model using the optical/UV emission, given that it is conceivable that a star of several solar mass was disrupted in the TDE. Instead, if we take the wind velocity as the escape velocity at $r_{\rm th}$ ($\sim2\times 10^3$ km/s), the amount of mass lost to the outflow is $<0.3 M_\odot$, which is less than half of the debris for a TDE of a solar-like star.  The main caveat to this calculation is our approximate treatment of photon diffusion in cases with $f_{\rm A} \ll 1$, which we hope to re-examine with a more detailed calculation in the future.

\subsubsection{Static Debris Layer}

Turning now to a quasi-static reprocessing layer, we approximate its radial density profile as a power-law, 
\begin{equation}
\rho(r)=\frac{K}{r^p}=\frac{M_{\rm s}(3-p)}{4\pi(r_o^{3-p}-r_i^{3-p})r^p},
\label{rho}
\end{equation}
where $K$ is a normalization constant, $M_{\rm s}$ is the layer mass if $f_{\rm A}=1$, and $p$ is expected to vary between 1.5 and 3 \citep{Coughlin+2014}. If $p=3$, $M_{\rm s}=4\pi K\ln(r_o/r_i)$. 

For the model involving a static layer of material, the trapping radius is not defined, as the gas has no bulk (outflow) motion. Observations can thus be characterized solely by the thermalization radius. We use the simplified Eq.~\ref{tau_eff} to described the effect of the optical depth at $r_{\rm th}$. The observed optical/UV black body luminosity can be also described by the diffusion approximation as shown in Eq.~\ref{dt}. By assuming $\frac{d (T^4)}{d r}\approx-\frac{T_{\rm bb}(t)^4}{r}$ at $r=r_{\rm th}$, we get  
\begin{equation}
K= (3\kappa_{\rm es})^{\frac{2p-3}{5}}\kappa_0^{\frac{-1-p}{5}}(16\pi\sigma_{\rm SB})^{\frac{2-3p}{5}}L_{\rm disc}(t)^{\frac{3p-2}{5}} T_{\rm bb}(t)^{\frac{23-17p}{10}}.
\end{equation}

The required mass of the layer at each epoch is,
\begin{equation}
M_{\rm{static}}= f_{\rm A}(t) M_{\rm s}.
\label{stat}
\end{equation}
We set the inner radius of the layer as $r_{\rm i} = 2r_{\rm t}$ (typically $\sim10^{13}$ cm for a $10^6-10^7 M_\odot$ BH), and $r_{\rm out} \sim 10^{15}$ 
$\rm{cm}$ \citep{Loeb+97,Coughlin+2014,Guillochon+14}.

\begin{figure}
\centering
\includegraphics[height=0.28\textheight]{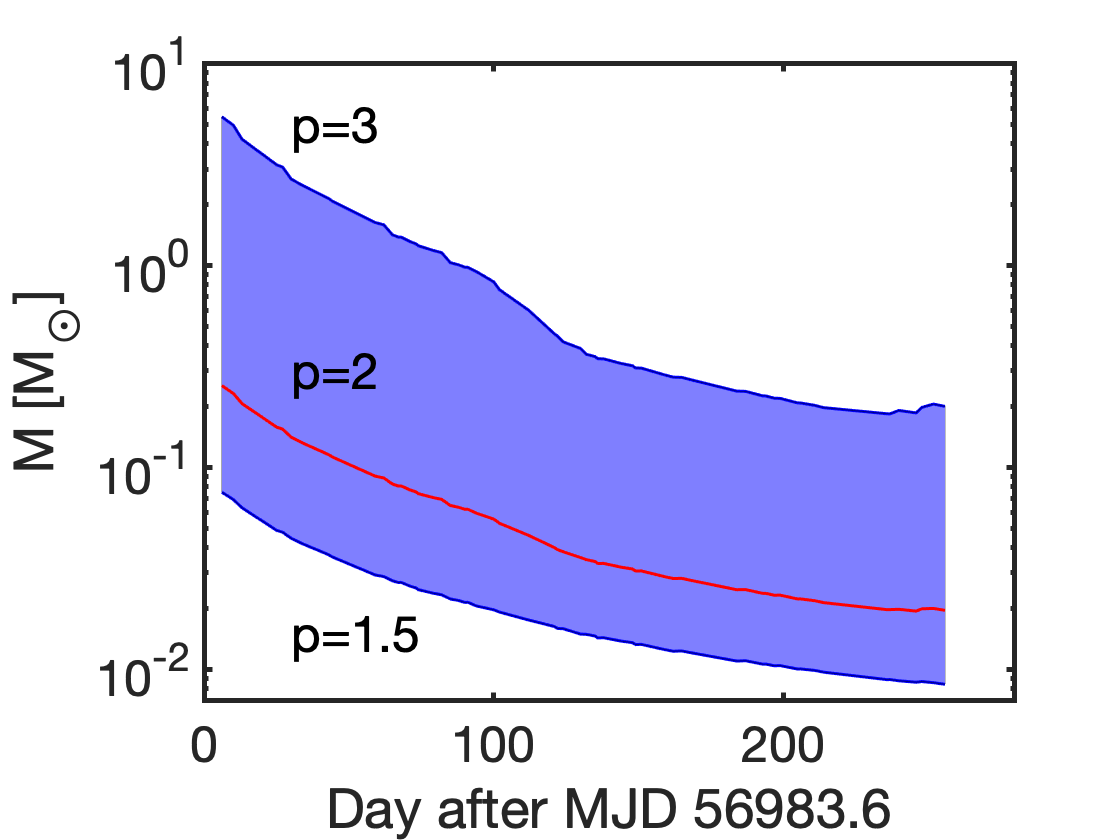}
\caption{Mass of the static layer as a function of time. Here, we set $r_{\rm out} = 1.7\times 10^{15}$ cm \citep{Loeb+97}, estimated by assuming constant ratio of gas pressure to total pressure for radiation supported layer. The mass in the static layer can be $\sim 0.5 M_\odot$,  indicating that the static layer model to explain the optical/UV emission is not ruled out given that it is not inconceivable that a solar mass star was disrupted in the TDE.}
\label{Fig:static}
\end{figure}

Figure \ref{Fig:static} shows the constraints on the mass of the gas for the static layer. The plot shows that the mass of the layer decreases with time, due to the quickly decreasing optical/UV luminosity. The mass also decreases with increasing $M_\bullet$, because a higher $M_\bullet$ results in a brighter $L_{\rm disc}$, which in turn makes the covering factor smaller. As the maximum mass in the static layer is $\sim 0.5 M_\odot$, we cannot rule the static layer model out. 

\section{Conclusions}
\label{Conclusions}
We have explored whether 
a TDE's optical/UV emission at late times---defined as times after the initial power-law decline of the light curve, and typically hundreds of days after discovery---can be described through quasi-thermal emission from a slim accretion disc. We examined how theoretical disc emission predictions 
are affected by the SMBH mass, SMBH spin, and an evolving outer disc radius. Applying slightly updated results (see \cite{Wen+22})
from our earlier slim disc modeling of 
ASASSN-14li's multi-epoch X-ray spectra (W20),
we 
tested if it is possible to fit the late-time \emph{Swift}
$u$-, $w1$-, $m2$-, and $w2$-band
luminosities
with only the host interstellar extinction and the time-dependent outer disc radius as free parameters. 
To better understand the \emph{early}-time optical/UV light curves, we 
explored constraints on two leading models explaining this emission:
(1) shock power arising from self-intersecting debris streams and (2) the reprocessing of obscured, inner-disc X-ray emission by an outflowing or static debris layer.

Our findings are:

\begin{enumerate}

\item The synthetic optical/UV slim disc luminosity depends on the inclination, the mass accretion rate, the SMBH mass and spin, and outer disc radius. We fixed the disc inclination 
to its best-fit value, and we fixed the mass accretion rate as a function of time to that found from the X-ray spectral fits. Expanding the outer disc edge increases the disc's emitting area and so increases the
light curve normalization. Increasing the SMBH mass 
(while holding the X-ray luminosity constant) decreases the disc temperature, shifting more of its total luminosity into longer wavelengths and increasing the optical/UV luminosity.  Increasing the spin (while holding the X-ray luminosity constant) increases the total radiative efficiency of the disc and thus modestly increases the optical/UV luminosity; this effect is generally weak except at the highest masses where the disc is much smaller in dimensionless gravitational radii.
Viscous spreading  
makes the light curves decay more slowly, but this effect weakens for larger $M_\bullet$, where spreading times are longer.

\item The optical/UV
luminosities across all four bands depend similarly on the black hole mass, black hole spin, and disc
outer radius.
Therefore, 
optical/UV colour (e.g., $u - w2$) is roughly constant with time and is 
insensitive to $M_\bullet$, $a_\bullet$, and the disc radius, as well as to the  accretion rate.  The one exception to this conclusion is if the disc spreads quickly, in which case colours may evolve. 

\item By including a starburst extinction model, allowing the disc outer radius to float, and using the best-fit $M_\bullet$, $a_\bullet$, disc inclination, and mass accretion rates
from our slim disc modeling of 
ASASSN-14li's X-ray spectra, we successfully fit all four optical/UV light curves at $t > 350$ days (see Figure \ref{Fig:lateUV}).  This further substantiates the finding of \citep{vanVelzen+19} that, contrary to simple time-dependent $\alpha$-disc models \citep{ShenMatzner14}, most TDE discs remain both thermally and viscously stable during periods of radiation-dominated, sub-Eddington accretion.

\item The application of the starburst galaxy dust extinction model provides the best and only good fit; the reduced $\chi^2$ is about 1.6 for 43 degrees of freedom and the fitted disc radius is consistent with 
our prior X-ray spectral fits.
This preferred extinction model is consistent with the fact that ASASSN-14li occurred in a post-starburst galaxy.

\item At late times, the fitted disc outer radius is about $2R_{\rm t}$ and, at the $1\sigma$ CL, is consistent with a constant value.  
This is significantly less viscous spreading than would be expected in simple models.  One relatively straightforward interpretation is that the ASASSN-14li inner disc lost large amounts of angular momentum, either due to preferential ejection directions during the circularization process \citep{BonnerotLu19} or in a magnetized wind \citep{BlandfordPayne82}.

\item
If ASASSN-14li's early optical/UV emission is powered by a shock arising from intersecting debris streams, the mass of the disrupted star 
is $M_\star > 1.4 M_\odot$ 
for $M_\bullet = 10^{6.5}$M$_\odot$, the low end of the
ASASSN-14li black hole mass range
($10^{6.5-7.1}$M$_\odot$; $1\sigma$ CL; W20).
For M$_\bullet >
10^{6.75}$M$_\odot$, 
the lower limit on progenitor mass becomes much more restrictive: 
$M_\star > 10 M_\odot$. 
Standard stellar initial mass functions produce relatively few stars at this high mass, and even those stars should no longer exist in ASASSN-14li's host galaxy, whose starburst ended $\sim$400 Myr ago \citep{French2017}.  Thus, while shock power passes the self-consistency checks we pose in \S \ref{sec:shocks}, it can only do so in a subset of the \{$M_\bullet, a_\bullet$\} region that is allowed by the X-ray data.

\item
It is also possible that ASASSN-14li's early optical/UV emission is generated 
through the reprocessing of X-ray emission
from the inner disc. 
We checked whether this reprocessing can be achieved for ASASSN-14li with a reasonable mass budget, and found that the gas mass required to explain the optical/UV luminosity can be lower than $\sim$0.5 $M_\odot$,
regardless of whether the reprocessing layer is static or outflowing in a wind.
Because this mass budget 
is consistent with a reasonable progenitor star mass ($\sim$1$M_{\odot}$ or less), the reprocessing model cannot be excluded as an explanation of the early optical/UV emission.  This result differs from some past work \citep{MatsumotoPiran20} in its consideration of a low covering fraction for wind reprocessing
and of a quasi-static reprocessing layer; both of these options reduce the required mass relative to a spherical outflow.

\end{enumerate}

Almost a decade after detection, ASASSN-14li remains a key laboratory for TDE physics.  By fitting its late-time optical/UV light curve with a sequence of slim disc models, we have further confirmed previous observational conclusions that late-time TDE optical/UV emission is dominated by a bare, relatively compact, thermally stable accretion disc \citep{vanVelzen+19}.  Somewhat surprisingly, this disc appears to have undergone little to no viscous spreading, which carries interesting implications for TDE disc physics.  

We have shown that the combination of early-time X-ray and optical/UV datasets allows for novel consistency tests on different hypotheses concerning a central question in TDE physics: what is the geometry and power source of the early-time optical/UV photosphere?  While in this case our consistency checks failed to falsify either the shock or reprocessing paradigms, they carried interesting implications for both, and may prove more decisive in future events.  Specifically, X-ray bright TDEs with low optical/UV luminosities may struggle to self-consistently satisfy the shock paradigm, while (as already explored in \citep{MatsumotoPiran20}), the most optically bright TDEs run into self-consistency problems when described by the reprocessing picture.

While the near future carries the exciting prospect of hundreds if not thousands of TDE detections by ongoing and upcoming time-domain surveys, the scarcity of followup resources means that most of these events will lack the high-quality, multiwavelength coverage of past benchmark TDEs like ASASSN-14li.  The unexpected persistence of late-time, slowly evolving disc emission shows that the ``first generation'' of TDEs may continue to yield scientific returns (and further surprises) long after they have become a minority of all TDEs.

\section*{Acknowledgements}

We thank the referee for the helpful comments. SW thanks the Department of Astrophysics/IMAPP at Radboud University and the Department of Astronomy/Steward Observatory at the University of Arizona for post-doctoral support. SW also thanks Dongdong Shi for his useful discussion on host extinction models.
NCS received financial support from the Israel Science Foundation (Individual Research Grant 2565/19) and the United States-Israel Binational Science Foundation (Grant No. 2019772). 
AIZ acknowledges support from NASA ADAP grant \#80NSSC21K0988. She also thanks the hospitality of the Columbia Astrophysics Laboratory at Columbia University, where some of this work was completed. 
Our calculations were carried out at UA on the El Gato and Ocelote supercomputers, which are supported by the National Science Foundation under Grant No.~1228509.

\section*{Data Availability}
The data underlying this article will be shared on reasonable request to the corresponding author.




\begin{appendix}

\section{Fitting the X-ray spectra with a slim disc model}
\label{app:decay}

We simultaneously refit the 10 epochs of X-ray spectra with the slim disc model of W21. The high-level fit assumptions are the same as in W20, i.e. all epochs have the same $M_\bullet$, $a_\bullet$ and $\theta$, but different absorption parameters $N_H$ and $\dot m$.

\begin{figure*}
\includegraphics[height=0.65\textheight]{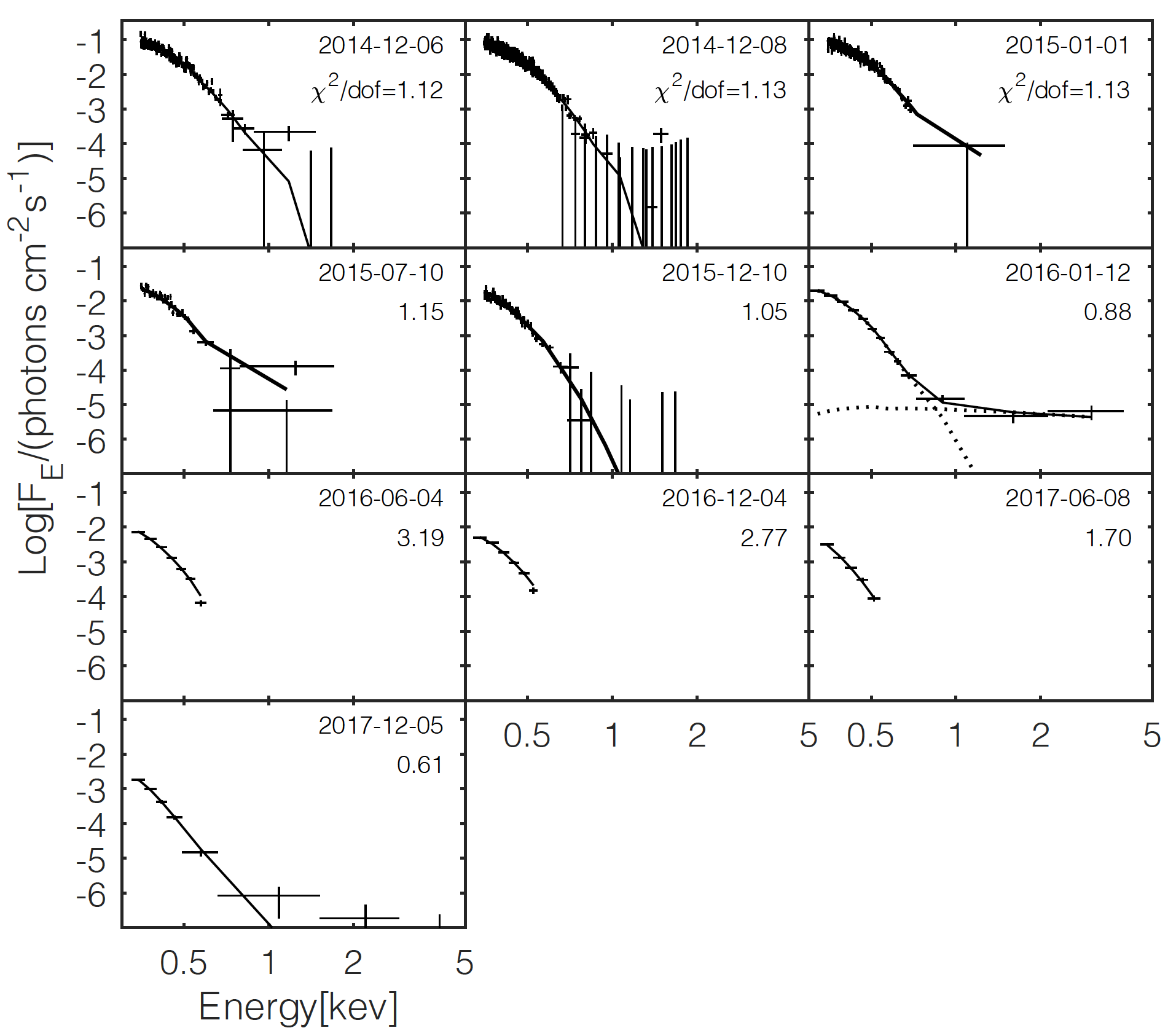}
\caption{Simultaneous slim disk fits to XMM--{\it Newton} spectra of ASASSN-14li. The first five spectra are obtained using the two RGS detectors (we only plot RGS1),
which provide data over the energy range 0.35--1.9 keV. The last five spectra are obtained using the 
pn detector, which is sensitive over 0.3--10 keV. All spectra are background-subtracted, and the data are binned so that there are at least 30 counts per bin. Each panel shows the best-fit model as a solid black line.  In nine of the epochs, only the quasi-thermal disk model is needed to fit observations. In Epoch 6, however, we need an additional power-law component to fit the hard emission (horizontal dotted line). The fitting results are in Table~\ref{Tab:Xray}. }
\label{14sp}
\end{figure*}

\begin{table}
\centering
\caption{Best-fit results for ASASSN-14li, with $M_\bullet=10^7M_{\odot}$ and $a_\bullet=0.998$. In our simultaneous fitting, the $M_\bullet$ and $a_\bullet$ are held fixed, inclination is required to be the same for all epochs, and other parameters float. The total $\chi^2_{dof}$ is $4367.49/3895=1.12$. The power-law parameters of epoch 6 are: $\Gamma=1.6\pm0.7$ and $A_{\rm pl}=8.1\pm2.7\times 10^{-6}$ $\rm {photons ~s^{-1}~cm^{-2}~keV^{-1}}$.}
\begin{tabular}{cccccc}
\hline
 Epoch  & $N_H$ $ [10^{20}{\rm cm}^{-2}]$& $\theta$ $ [^\circ]$  & $\dot m$ $ [Edd]$  &  $\chi^2/d.o.f$ \\
\hline
           1   &$5.4\pm0.3$   &$76\pm3$  &$1.09\pm0.05$   
           &630.1/563 \\
           2   & $5.1\pm0.2$   & ...  & $1.03\pm0.04$ 
           & 2322.33/2057     \\
           3   &$5.2\pm0.3$  &... &$1.06\pm0.05$ 
           &591.18/521 \\
           4  &$5.4\pm0.4$   &...  &$0.44\pm0.01$
           &  178.93/155      \\ 
           5  &$4.1\pm0.3$ &... &$0.35\pm0.01$ 
           & 601.6/572     \\
           6   &  $4.4\pm0.3$ &...  &$0.335\pm0.005$   & 7.02/8   \\ 
            7  & $4.1\pm0.4$ & ...  &$0.227\pm0.004$   & 15.96/5    \\ 
            8  &$5.1\pm0.5$ & ...   &$0.206\pm0.006$     &   11.1/4    \\
           9   &$2.6\pm0.8$ & ... &  $0.150\pm0.006$  & 5.09/3  \\ 
            10  &$3.5\pm0.8$& ... &$0.130\pm0.005$  & 4.27/7    \\
\hline
\end{tabular}
\label{Tab:Xray}
\end{table}

Figure \ref{14sp} shows the best fit to the 10 epochs of spectra. As shown in the figure, our model fits the spectra well, with the total reduced $\chi^2$ being 1.12. The fitting parameters are listed in Table \ref{Tab:Xray}. The total $\chi^2$ is about 1 smaller than that of 4368.42 in W20. The fitted $N_H$ and $\theta$ are consistent with those in W20 at the $1\sigma$ CL. However, the fitted $\dot m $ here are about $20\%$ higher than those in W20 (note that $\dot m$ presented here are not corrected for radiative efficiency). The slim disc model of W21 employed here differs from the earlier model in W20 by including the effect of angular momentum loss by radiation. This effect will make the disc cooler, resulting in a higher fitted $\dot m$. However, both the W20 and W21 models produce similar constraints on $M_\bullet$ and $a_\bullet$.

\begin{figure}
\includegraphics[height=.28\textheight]{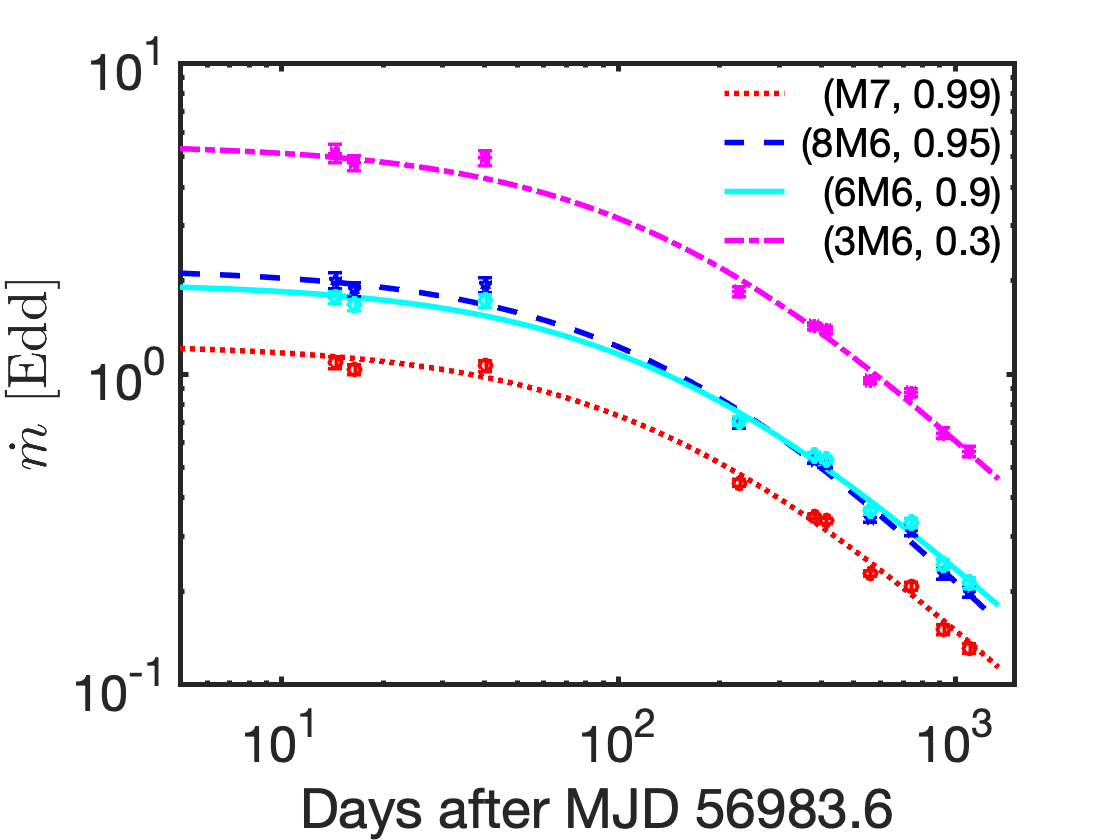}
\caption{Dimensionless accretion rates $\dot{m}$ for four $\{M_\bullet$, $a_\bullet\}$ pairs obtained from X-ray spectral fitting, along with the corresponding best fit power-law $\dot{m}(t)$ functions. The equations are $\dot m_t= 231(t+150)^{-1.04\pm0.02}$, $\dot m_t= 658(t+150)^{-1.14\pm0.03}$, $\dot m_t= 361(t+150)^{-1.04\pm0.02}$, and $\dot m_t= 1227(t+150)^{-1.08\pm0.02}$, from top to the bottom of the legend, respectively.
}
\label{Fig:mdot}
\end{figure}

In order to constrain the decay of $\dot m$ for each pair of $M_\bullet$ and $a_\bullet$ that lie within the $1\sigma$ contour of Fig. 6 of W20, we calculate the best-fit and $1\sigma$ error bars on the $\dot m$ \footnote{Here we use slim disc model {\tt slimd} \citep{Wen+22} to do the fitting. We note that {\tt slimd} is the tabulated version of W21.}. We then fit the $\dot m$ with a power law function $\dot m_t=A(t+150)^{-n}$, where $A$ and $n$ are free parameters to be fitted. Figure \ref{Fig:mdot} shows the power law fitting result for 4 pairs of $M_\bullet$ and $a_\bullet$. We can see that the decay in the accretion rate can be well fit by a power law function, especially for the late time epochs.

\section{Disc spreading}
\label{app:spread}
In this Appendix, we derive a toy model for the time evolution of a viscously spreading disc, to derive approximate expectations for the evolution of its outer radius $R_{\rm out}$.  We assume the disc possesses a surface mass density profile $\Sigma(R) = \Sigma_{\rm out} (R/R_{\rm out})^{-\Gamma}$ for all radii $R_{\rm ISCO} \le R \le R_{\rm out}$, and has no mass elsewhere.  We assume that the disc begins with an initial mass $M_\star/2$, or in other words the entire bound mass of the disrupted star.  This approach, similar to that in \citep{vanVelzen+19} and \citep{Mummery&Balbus20}, will be quite inaccurate at early times, but should be reasonably accurate at late times when $t_{\rm visc} \gg t_{\rm fall}$.  The instantaneous disc mass $M_{\rm d} = M_\star/2 - \Delta M$, where $\Delta M$ is the total mass accreted so far.  The instantaneous disc angular momentum $L_{\rm d} = (M_{\rm d}+\Delta M)J_{\rm circ} - \Delta L$, where $J_{\rm circ}=\sqrt{2GM_\bullet R_{\rm circ}}$ is the specific angular momentum of the disrupted star and $\Delta L = J_{\rm ISCO} \Delta M$ is the angular momentum lost into the event horizon through accretion.  As this is just a toy model, we crudely assume that $J_{\rm ISCO} = \sqrt{GM_\bullet R_{\rm ISCO}}$, and that $R_{\rm ISCO} = 6R_{\rm g}$.

By integrating over the disc size, we find that $\Sigma_{\rm out} \approx (2-\Gamma)M_{\rm d}/(2\pi R_{\rm out}^2)$ so long as $\Gamma < 2$.  Likewise, we find that $L_{\rm d} \approx 2\pi (5/2-\Gamma)^{-1}\Sigma_{\rm out}\sqrt{GM_\bullet} R_{\rm out}^{5/2}$ so long as $\Gamma < 5/2$.  In both cases the approximate equalities come from assuming that $R_{\rm out} \gg R_{\rm ISCO}$, and the assumptions on $\Gamma$ are likely valid (for example, in the radiation-dominated regime of a Shakura-Sunyaev disc, $\Gamma = -3/2$).

We now combine these approximate equalities with our earlier statements of mass and angular momentum conservation to solve for the expanding outer edge of the spreading disc:
\begin{equation}
    R_{\rm out} = \left( \frac{5/2-\Gamma}{2-\Gamma} \right)^2 \left(\frac{M_\star \sqrt{R_{\rm circ}/2}-\Delta M \sqrt{R_{\rm ISCO}}}{M_\star/2 - \Delta M} \right)^2.
    \label{app:rout}
\end{equation}
As an example, consider $R_{\rm ISCO} = 6R_{\rm g}$ and $R_{\rm circ} = 100 R_{\rm g}$.  This equation then implies that if the disc has lost $80\%$ of its initial mass ($\Delta M = 0.8 M_\star /2$), then $R_{\rm out}$ should expand by a factor $\approx 19$.  As we see in Fig. \ref{Fig:Rout}, no evidence for significant disc spreading is seen.  There are a few plausible explanations for this.  First, it is possible that very little of the disc mass has actually accreted onto the MBH.  In the above example, if $\Delta M = 0.2 M_\star / 2$, then $R_{\rm out}$ will only grow by a factor $\approx 1.4$, which is marginally consistent with Fig. \ref{Fig:Rout}.

A second possibility is that our assumption that $\Delta L = J_{\rm ISCO} \Delta M$ is incorrect, and that there is a more important additional source of angular momentum loss.  For example, if large amounts of angular momentum are lost in a magnetized wind \citep{Waters2018} (as is seen in MHD simulations of super-Eddington accretion discs), this may dominate $\Delta L$ and limit the disc's ability to spread.  More speculatively, the early time shocks between the disc and returning debris streams may preferentially eject high-angular momentum material from the system, starving the disc of its angular momentum budget.

A third possibility is that $R_{\rm circ} \approx R_{\rm ISCO}$, or in other words that this was a rare, high-$\beta$ TDE.  However, a significant degree of fine-tuning is needed to eliminate observable changes in $R_{\rm out}$ through this explanation.

\end{appendix}

\label{lastpage}
\end{document}